\documentclass[reprint,nobibnotes,aps,prb,amsbsy,superscriptaddress,floatfix]{revtex4-2}

\usepackage{epsfig,grffile}
\usepackage{graphicx,graphics}
\usepackage{caption}
\usepackage{ragged2e}
\usepackage[justification=justified,singlelinecheck=false]{caption}
\usepackage{subcaption}
\usepackage[table]{xcolor}
\usepackage{epstopdf}
\usepackage{amsmath,amssymb,amsfonts,stmaryrd,amsthm,mathrsfs}
\usepackage{latexsym,verbatim}
\usepackage{bm}
\usepackage{bbold}
\usepackage{color}
\usepackage{ulem}
\usepackage[breaklinks=false,colorlinks,citecolor=blue,linkcolor=blue,urlcolor=blue]{hyperref}
\usepackage[capitalise]{cleveref}
\usepackage[shortlabels]{enumitem}
\usepackage[abs]{overpic}
\usepackage{textcomp} 
\usepackage{mathtools}
\usepackage{braket}
\usepackage{comment}
\usepackage{tcolorbox}
\usepackage{dutchcal}
\usepackage{soul}    
\usepackage{xspace}
\usepackage{acronym}
\usepackage{orcidlink}
\usepackage{xr} 
\usepackage[version=4]{mhchem}

\newcommand*{\fq}{FQ\xspace}
\newcommand*{\fqfmu}{FQF$\mu$\xspace}
\newcommand*{\wfq}{$\omega$FQ\xspace}
\newcommand*{\wfqfmu}{$\omega$FQF$\mu$\xspace}
\newcommand*{\tqq}{T^\mathrm{qq}}
\newcommand*{\tqmu}{T^{\mathrm{q}\mu}}
\newcommand*{\tmuq}{T^{\mu\mathrm{q}}}
\newcommand*{\tmumu}{T^{\mu\mu}}
\newcommand{\absu}[1]{\left|{#1}\right|}

\DeclareMathOperator{\tr}{tr}

\DeclareMathOperator{\diag}{diag}

\usepackage{physics}
\usepackage{tensor}
\usepackage{abbrevs}
\usepackage[capitalise]{cleveref}

\DeclareFontFamily{U}{mathx}{\hyphenchar\font45}
\DeclareFontShape{U}{mathx}{m}{n}{
      <5> <6> <7> <8> <9> <10>
      <10.95> <12> <14.4> <17.28> <20.74> <24.88>
      mathx10
      }{}
\DeclareSymbolFont{mathx}{U}{mathx}{m}{n}
\DeclareFontSubstitution{U}{mathx}{m}{n}
\DeclareMathAccent{\widecheck}{0}{mathx}{"71}
\DeclareMathAccent{\wideparen}{0}{mathx}{"75}

\begin{document}
 
\title{Quantum Dynamics of Dissipative Polarizable Media}

\author{Frank Ernesto Quintela Rodriguez\,\orcidlink{0000-0002-9475-2267}}
\email[Email: ]{frank.quintelarodriguez@sns.it}
\affiliation{Scuola Normale Superiore, Piazza dei Cavalieri, 7, Pisa, 56126, Italy}

\author{Piero Lafiosca\, \orcidlink{0000-0002-3967-0736}}
\affiliation{Scuola Normale Superiore, Piazza dei Cavalieri, 7, Pisa, 56126, Italy}

\author{Tommaso Giovannini\,\orcidlink{0000-0002-5637-2853}}
\affiliation{Department of Physics, University of Rome Tor Vergata, Via della Ricerca Scientifica 1, 00133, Rome, Italy}

\author{Chiara Cappelli\,\orcidlink{0000-0002-4872-4505}}
\email[Email: ]{chiara.cappelli@sns.it}
\affiliation{Scuola Normale Superiore, Piazza dei Cavalieri, 7, Pisa, 56126, Italy}
\affiliation{IMT School for Advanced Studies Lucca, Piazza San Francesco 19, Lucca, 55100, Italy}

\begin{abstract} 
    Classical polarizable approaches have become the gold standard for simulating complex systems and processes in the condensed phase. 
    These methods describe intrinsically dissipative polarizable media, requiring a formal definition within the framework of open quantum systems. 
    We present a Hamiltonian formulation for the quantum dynamics of polarizable sources based on a generalized theory of the damped harmonic oscillator, using pseudo-boson theory to characterize their coherent state dynamics. 
    We then apply our theory to the study of the optical response of two plasmonic systems.
    Furthermore, by exploiting the phase space formulation of quantum mechanics and the integrability of quadratic Hamiltonians, we derive a self-consistent relation for the emitted electric field of the polarizable medium under the semiclassical approximation, based on exact formulas for medium polarization.
    Finally, we derive the master equation describing the open dynamics of a quantum system interacting with the quantum polarizable medium, along with analytical expressions for correlation functions calculated over arbitrary Gaussian states.
\end{abstract}

\date{\today}

\maketitle

\section{Introduction}

Simulating the molecular properties of complex systems under the effects of external electromagnetic fields is a challenging topic in theoretical chemistry and condensed matter physics. 
Methods based on \ac{QM} become computationally prohibitive for systems larger than a thousand atoms, limiting their applicability to real systems~\cite{https://doi.org/10.1002/wcms.1290,doi:10.1098/rsta.2002.0993,sisto2017atomistic}.
A successful solution is multiscale modeling, which divides the system into a target and an environment~\cite{nobel2013,warshel1976theoretical,field1990combined,senn2009qm,Horstemeyer2010,weinan2011principles}. 
The target, representing the relevant degrees of freedom, is modeled using an accurate quantum method, while the environment is modeled with a simpler approach focusing on its impact on the target dynamics~\cite{Reuter2000,https://doi.org/10.1002/anie.200802019}.
For instance, in \ac{CPM} methods, the quantum dynamics is approximated by considering polarization effects to external sources as the key factor for accurately describing most physicochemical phenomena~\cite{giovannini2022we}.

\ac{CPM} methods have become the golden standard for simulating complex systems and processes in the condensed phase thanks to the good compromise between accuracy and computational cost~\cite{bondanza2020polarizable}.
They have been amply applied in various research fields~\cite{tomasi2005quantum,senn2009qm,mennucci2019multiscale,lin2007qm}, including the simulation of the structural dynamics of large biomolecules~\cite{baker2015polarizable}, chemical reactivity~\cite{senftle2016reaxff} and optical properties of plasmonic materials~\cite{coccia2020hybrid,draine1994discrete,de2002retarded,jensen2009atomistic,zakomirnyi2019extended,giovannini2019classical,giovannini2022we,garcia2014graphene,giovannini2020graphene}. 
Furthermore, they have been coupled with \ac{QM} methods in a multiscale framework for simulating structural and spectral properties of solvated systems~\cite{loco2021atomistic,giovannini2023continuum}, complex biomolecular structures~\cite{senn2009qm,bondanza2020polarizable} and enhanced spectral signals of nanostructured plasmonic materials~\cite{coccia2020hybrid,payton2014hybrid,lafiosca2023qm}. 
Moreover, the investigation of plasmon-molecule hybrid states occurring in molecular systems strongly coupled to plasmonic nanocavities using \ac{CPM} has been recently proposed~\cite{fregoni2021strong}.

In all applications involving \ac{CPM}s, energy transfer to an external system -- whether it be a dissipative bath or a \ac{QM} system in multiscale approaches -- is inherently modeled.
To formally describe this process, it is essential to refer to the theory of open quantum systems, a key aspect that has generally been overlooked in the literature~\cite{guido2020open}.
In this work, we establish the theoretical foundations for treating any \ac{CPM} within the framework of open quantum systems. Our theoretical framework bridges traditional \ac{CPM} applications with quantum theory, expanding their utility for capturing complex \ac{QM} processes. 
Our approach thus lays the groundwork for studying purely quantum phenomena, such as energy and information transfer, entanglement, and decoherence, using models commonly applied in condensed matter physics simulations.

A generic \ac{CPM} is represented by a set of electrostatic variables consisting of: 
\begin{itemize}
    \item Time-dependent polarization sources ${u_\alpha (t)}$ indexed by ${ \alpha \in \{ 1, \dots, n\} }$, representing electric charges, dipoles or multipoles.
    \item A set of generalized coordinates ${R_\alpha }$, associated with the polarization sources, which is assumed to be fixed.  When the sources ${u_\alpha (t)}$ represent charges, then ${R_\alpha \in \mathbb R^3 }$ correspond to their spatial coordinate, and is given in components by ${R_{j \alpha}}$, with ${j=1,2,3}$.
    \item Covariance matrices ${\Sigma_\alpha}$ describing the spatial dispersion of the polarization sources. When the sources ${u_\alpha(t)}$ represent charges, the coordinates and covariances parametrizes a Gaussian distribution $G (r;\, R_\alpha, \Sigma_\alpha)$, where ${\Sigma_\alpha \in \mathbb R^{3\times 3}}$.
\end{itemize}
Table \ref{tab:models} specifies the electrostatic variables and polarization sources for widely used \ac{CPM} approaches, which are further described in the Supplemental Material~\footnote{ See Supplemental Material (Section I) at \url{http://} for a definition of the acronyms and further details on the diverse models} (see also references~\cite{rick1994dynamical,rick1995fluctuating,thole1981molecular,ohno1964some,mayer2007formulation,giovannini2019polarizable,thompson1996qm,olsen2010excited,jensen2003discrete,ponder2010current,pipolo2014cavity,klamt1993cosmo,cances1997new,tomasi2005quantum,giovannini2019classical,giovannini2020graphene,lafiosca2024real,lafiosca2021going,giovannini2022we,jensen2009atomistic,pipolo2016real,dall2020real} therein).
Within these approaches, each model establishes a relationship between the atomic/molecular structure and the polarizable sources, meaning there is no inherent one-to-one correspondence between the system's atomic structure and the number and position of the ${u_\alpha (t)}$ variables.
\begin{table}[!ht]
    \centering
    \begin{tabular}{ccccc}
    \hline
    \hline
    Model & Electrostatic & Polarization & Decoherence & Refs.\\
    \hline
    FQ & - & $q$ & - & \cite{rick1994dynamical} \\
    FQF$\mu$ & - & $q,\mu$ & - & \cite{giovannini2019polarizable} \\
    IPD/MMPol & $q$ & $\mu$& - & \cite{thole1981molecular} \\
    PE & $q$, $\mu$, $\Theta$, $\Xi$ & $\mu$ & - & \cite{olsen2010excited} \\
    AMOEBA & $q$, $\mu$, $\Theta$& $\mu$& - & \cite{ponder2010current}\\
    DRF & $q$& $\mu$& - & \cite{jensen2003discrete}\\
    COSMO & - & $\sigma$ & - & \cite{klamt1993cosmo}\\
    IEFPCM & - & $\sigma$ & - & \cite{cances1997new}\\
    $\omega$FQ & -& $q$& yes & \cite{giovannini2019classical}\\
    $\omega$FQF$\mu$ & - & $q$, $\mu$ & yes & \cite{giovannini2022we}\\
    DIM & $q$ & $\mu$ & yes & \cite{jensen2009atomistic} \\
    BEM & -& $\sigma$ & yes & \cite{de2002retarded}\\
    \hline
    \hline
    \end{tabular}
    \caption{ \justifying List of \ac{CPM} models exploited in condensed matter simulations. They are specified in terms of permanent electrostatic quantities (charges $q$, dipoles $\mu$, quadrupoles $\Theta$, and octupoles $\Xi$) and polarization sources $u_
    \alpha$ (charges $q$, dipoles $\mu$, surface charge density $\sigma$). 
    For purely dissipative approaches, a decoherence term is considered. }
    \label{tab:models}
\end{table}

While frozen in the coordinate ${R_\alpha }$, the polarization sources ${u_\alpha (t)}$ are driven in time by an electric field ${f_\alpha (t)}$, which can be generated either by the \ac{QM} target or an external source.  
The interaction strength between the polarization sources is represented by the matrix ${K := K( \{ R_\alpha\}, \{ \Sigma_\alpha \} ) }$, ${ K \in \mathbb R^{n\times n}}$, which depends on the sources' spatial distribution and may also include an effective description of quantum tunneling~\cite{giovannini2019classical,giovannini2020graphene}.
Decoherence mechanisms are effectively modeled by a diagonal matrix ${ \Gamma \geq 0,\, \Gamma \in \mathbb R^{n\times n} }$~\cite{mennucci2019multiscale,giovannini2019classical,jensen2009atomistic}, embodying the assumption of a memory-less dissipation.
From a formal point of view, the \ac{CPM} consists of a ${(4n+2)}$-tuple ${ \big\{ u_\alpha (t), R_\alpha, \Sigma_\alpha, \Gamma, K, f_\alpha (t) \big\} }$ verifying the \mbox{non-homogeneous} \mbox{second-order} linear differential equation
\begin{align} \label{m_185}
    & \ddot{u}(t) + 2\Gamma\dot{u}(t) + K u(t) + f(t) = 0 ,
\end{align}
where we have used a vector notation for polarization sources.
In what follows, we will allow $\Gamma$ to be an arbitrary matrix, with the aim of extending its role in \cref{m_185} to account for the Lorentz interaction mechanism~\cite{Ghosh_2019,Ghosh21}, i.e., the effect of a uniform magnetic field.
Although the above formulation emphasizes the atomistic nature of the polarization sources~\cite{loco2021atomistic}, a common theoretical framework can be found with continuum descriptions~\cite{nottoli2020general,klamt1993cosmo,cances1997new}.

Here, we aim to formulate the dynamics of \ac{CPM}, as described by \cref{m_185}, in terms of a quadratic Hamiltonian.
The simplest example is obtained by setting ${\Gamma=0}$, resulting in the following Hamiltonian: 
\begin{align} \label{m_147}
    &  H (t) = \frac{1}{2} \pi^\mathrm T A \pi  + \frac{1}{2} u^\mathrm T A^{-1} K u + u^\mathrm T A^{-1} f(t),
\end{align}
where $A$ is the symmetric, invertible matrix that achieves the similarity transformation ${K = A K^\mathrm T A^{-1}}$~\cite{taussky1959similarity,horn2012matrix}. 
In general, $A$ can be obtained from the Jordan normal form decomposition of $K$, as will be discussed in detail later.
When ${K = K ^\mathrm T}$, we trivially have ${A= I}$, which occurs for example in the so-called Fluctuating Charge (FQ) model~\cite{cappelli2016integrated} described in the Supplemental Material~\footnote{ See Supplemental Material (Section I A 1) at \url{http://} for a derivation of the FQ model.} (see also references~\cite{rick1994dynamical,rick1995fluctuating,thole1981molecular,ohno1964some,mayer2007formulation,giovannini2019polarizable} therein).
The difficulties in finding a Hamiltonian for \cref{m_185} arise from the need to account simultaneously for the matrices $\Gamma$ and $K$. 
This effort is illustrated by several approaches, including the time-dependent Caldirola-Kanai Hamiltonian~\cite{caldirola83,Kanai48,Kahng87}, the expanding coordinates Hamiltonian derived in the Supplemental Material~\footnote{ See Supplemental Material (Section II) at \url{http://} for a derivation of the expanding coordinates Hamiltonian.} 
(see also references~\cite{goldstein2002classical,giovannini2019classical,taussky1959similarity} therein),
Bateman's Hamiltonian, which governs the global dynamics of the system coupled to an auxiliary (dual) one~\cite{bateman1931dissipative,10.1063/5.0039248}, and the canonical formulation of balanced loss and gain systems~\cite{Ghosh_2019,Ghosh21}, among others. 
Alternatively, following the Caldeira-Leggett model~\cite{PhysRevLett.46.211}, a Hamiltonian can be derived by coupling the system degrees of freedom ${u_\alpha (t)}$ to a large (${n \to \infty }$) bath of harmonic oscillators, which, assuming an Ohmic spectral density, allows us to recover \cref{m_185} plus an additional noise term reflecting the fluctuation-dissipation relations characterizing the bath~\cite{weiss2012quantum}.
The relation between this and some of the above approaches is discussed in Refs.~\cite{PhysRevA.49.592,PhysRevA.51.1845,schuch2018quantum,Schuch2015,Schuch17}.

For the differential equation in \cref{m_185}, all of the aforementioned approaches provide a quadratic Hamiltonian, each of which may be more or less convenient depending on the specific problem at hand—specifically, the trade-off between time-dependent approaches versus time-independent formulations of higher dimensionality.
Here, we propose a different Hamiltonian approach, which adopts a method similar to Bateman’s, aimed at establishing a finite, time-independent Hamiltonian (if ${f(t)=0}$) that governs the global dynamics of the system coupled to auxiliary degrees of freedom. 
Unlike the Caldeira-Leggett model, this method does not require taking the limit ${n\to \infty}$ in the auxiliary degrees of freedom, nor does it require the inclusion of a noise term in the dynamical equation, which may be undesired in a general modeling framework that is not constrained by the fluctuation-dissipation theorem. 
However, unlike Bateman's approach, our formulation will not assume prior information about the auxiliary variables; the dynamical information of the global system will be fully specified by \cref{m_185}.

The paper is organized as follows:\\
\Cref{m_272}
presents the theoretical foundations of the paper. First, we show how \cref{m_185} can be integrated in an expanded space, where a time-independent matrix ${\sqrt \mathcal K}$ encapsulates the spectral information of the dynamics. 
Second, a quadratic Hamiltonian is proposed and quantized under the criterion that its mean dynamics coincide with \cref{m_185}, thereby defining the \ac{QPM}. The Hamiltonian is shown to provide a constant of motion (for ${f(t)=0}$) equal to zero for all times, consistent with the system’s long-term energetics. 
Third, the Hamiltonian is transformed to a diagonal form in terms of pseudo-boson operators, allowing us to express the dynamics over the set of Bi-Coherent states.

\Cref{m_244}
applies our theoretical results to the analysis of the polarization signal measurement. First, we derive the formulas for decomposing the polarization in terms of eigenvalues and eigenvectors. 
Second, by exploiting the spectral structure, we propose practical filtering methods to reproduce the polarization with a reduced number of eigenpairs.

\Cref{m_273} presents the calculation of the semi-classical electric field emitted by a polarizable medium following an electric excitation. 
First, the \ac{QPM} dynamics is formulated in phase space, providing a “classical” picture of the dynamics as well as tools for the analytical calculation of expectation values over arbitrary \ac{GS}, including the thermal state. 
Second, assuming the \ac{QPM} interacts with a classical field, a self-consistent equation is derived for the emitted electric field in frequency-momentum space. 
Third, an explicit solution is found perturbatively, and its properties are discussed with a view toward possible applications for electric field enhancement in nanoplasmonic materials.

\Cref{m_274}
 formulates the open quantum dynamics of a \ac{QM} system within the quantum polarizable medium (\ac{QPM}). First, building on the results for classical electrostatic fields in \cref{m_273}, we derive the quantum interaction between the \ac{QM} system and the \ac{QPM}, including the auxiliary variables, which are used to define the quantum master equation. 
 Second, we calculate the \ac{QPM} correlation functions for arbitrary initial \ac{GS}, focusing particularly on thermal states.

\Cref{m_275} presents the conclusions of our work, followed by acknowledgments for financial support and contributions from our collaborators.

\section{Generalized theory of the damped harmonic oscillator}  \label{m_272}

\subsection{Classical Dynamics} \label{m_356}

Introducing in \cref{m_185} the auxiliary variable ${v = \dot u }$~\cite{10.1115/1.4011828,https://doi.org/10.1002/eqe.4290090205,Sanders2022} and taking its first and second derivatives, we obtain the following equations:
\begin{align} \label{m_177}
    & \ddot u + K u + 2 \Gamma v + f(t) = 0 \\
    & \ddot v + ( K  - (2 \Gamma )^2 ) v - 2 \Gamma K u  + (\dot f(t) - 2 \Gamma f(t) ) = 0,
\end{align}
where we are implicitly indicating the time dependence of the variables ${u \equiv u (t)}$ and ${v \equiv v (t)}$, and 
${v(t)}$ is subject to the initial conditions ${(v(0),\, \dot v(0)) = ( \dot u(0), -K u (0) - 2 \Gamma \dot u (0) )}$.
Defining the vector ${x := \begin{bmatrix} u & v \end{bmatrix}^\mathrm T }$ and 
\begin{align} \label{m_178}
    & \mathcal K := \begin{bmatrix} K & 2\Gamma  \\ - 2 \Gamma K  &  K - (2 \Gamma )^2 \end{bmatrix} \\
\label{m_203}
    & F(t) := \begin{bmatrix} f(t) &  \dot f(t) - 2 \Gamma f(t) \end{bmatrix}^\mathrm T,
\end{align}
results the velocity independent equation 
\begin{subequations} \label{m_179} 
\begin{align} \label{m_296}
    & \ddot x + \mathcal K x + F(t) = 0\\
    \label{m_288}
    & x(0) = \begin{bmatrix} u(0) & \dot u(0) \end{bmatrix}^\mathrm T \\
    \label{m_351}
    & \dot x(0) = \begin{bmatrix} \dot u(0) & - K u(0) - 2\Gamma \dot u (0) \end{bmatrix}^\mathrm T.
\end{align}
\end{subequations}
Fixing the initial conditions is of utmost importance, and it can be easily verified numerically that, for the one-dimensional oscillator and reasonable model parameters, overlooking the conditions \cref{m_288,m_351} may lead to unphysical behavior or ill-defined evolutions. 
In other words, \cref{m_185} and \cref{m_296} are equivalent only if \cref{m_288,m_351} are satisfied. 

The homogeneous solution is constructed from linear combinations of the exponential functions  
\begin{align} \label{m_215}
    & \exp{\pm i \sqrt {\mathcal K}t},\quad \sqrt{\mathcal K} = i \begin{bmatrix} 0 & - I \\  K  &  2\Gamma \end{bmatrix} ,
\end{align}
with the coefficients determined from the initial conditions \cref{m_288,m_351}.
Note that, in the construction of the matrix ${\sqrt \mathcal K}$, no assumptions need to be made about the matrices ${K}$ and ${\Gamma}$; i.e., they can be non-real, non-invertible or even non-diagonalizable.
In what follows, we will focus on the spectral features of such solutions.

Let ${Sp \big\{ \sqrt\mathcal K \big\} }$ denote the set of eigenvalues of ${\sqrt \mathcal K}$. 
These eigenvalues can be determined from the zeros of the characteristic polynomial ${p_{\sqrt \mathcal K} (\lambda) = \det[\sqrt \mathcal K - \lambda I]}$, where ${I}$ the identity matrix in ${\mathbb R^{2n \times 2n} }$, together with the formula for the determinant of a block matrix:
\begin{align} \label{m_325}
    & \det \begin{bmatrix} A & B \\  C  &  D \end{bmatrix} = \det A \det (D - C A ^{-1} B),
\end{align}
provided that ${A}$ is invertible.
Combining these expressions, we can show that 
\begin{align} \label{m_324}
    & Sp \big \{ - \sqrt \mathcal K \big\} \subseteq \big\{ \omega : \det(\omega^2 + 2 i \omega \Gamma - K ) = 0 \big\} \cup \{0\}.
\end{align}
This result plays a central role in our analysis, serving to characterize the spectral properties of the time-evolved polarizable sources ${u (t)}$ in response to a delta-like pulse--- an aspect we will explore further in \cref{m_244}.\\
\indent The importance of the matrix ${\sqrt \mathcal K}$ was already recognized in Ref.~\cite{Ashida02072020}, and it was recently derived from a Lagrangian formulation in Ref.~\cite{krovi2024quantumalgorithmssimulatequadratic}.
However, to our knowledge, the spectral equivalence between ${-\sqrt \mathcal K}$ and the dynamical equation \cref{m_185} is a novel result and represents the first main outcome of our formulation.

\subsection{Hamiltonian Formulation}

In the extended variables system, \cref{m_179}, a Hamiltonian can be provided (similarly to \cref{m_147}) by the function  
\begin{subequations} \label{m_180}
\begin{align} \label{m_320}
    & H (t) = H_0 + H_1 (t) \\
    \label{m_321}
    & H_0 = \frac{1}{2} \pi^\mathrm T A \pi  + \frac{1}{2} x^\mathrm T A^{-1} \mathcal K x \\
    \label{m_322}
    & H_1 (t) = x^\mathrm T A^{-1} F(t)
\end{align}
\end{subequations}
where ${H_1(t)}$ isolates the time-dependent contribution.
Here ${A}$ is the symmetric and invertible matrix yielding the similarity transformation~\cite{taussky1959similarity,Kaczorek_2005}
\begin{align} \label{m_181}
    & \mathcal K = A \mathcal K^\mathrm T A^{-1} . 
\end{align}
It can be determined from the Jordan normal form decomposition of ${\mathcal K}$~\cite{Kaczorek_2005}, 
\begin{subequations} \label{m_225}
\begin{align} \label{m_221}
    & \mathcal K = P_1 J_\mathcal K P_1 ^{-1} ,\quad J_\mathcal K = \diag \begin{bmatrix} J_1 & J_2 & \dots & J_s \end{bmatrix} \\
    \label{m_222}
    & A := P_1 P_2 P_1 ^\mathrm T,\quad P_2 = \diag\, [ P_{2,1}\,  P_{2,2}\, \dots\, P_{2,s} ] \\
    \label{m_223}
    & P_2 ^\mathrm T = P_2 ^{-1} = P_2 ,
\end{align}
\end{subequations}
where the ${J_s}$ are \textit{Jordan blocks}, i.e., matrices of the form
\begin{align} \label{m_367}
    & J_s \coloneqq \begin{bmatrix} \lambda & 1 & 0 & \dots & 0  
                                   \\ 0 & \lambda & 1 & \dots & 0 
                                   \\ \vdots & \vdots & \ddots & \ddots & \vdots 
                                   \\ 0 & 0 & \dots  & \lambda & 1 
                                   \\ 0 & 0 & \dots & 0 & \lambda
    \end{bmatrix},
\end{align}
where the eigenvalue ${\lambda}$ corresponding to different Jordan blocks are not necessarily distinct.
The geometric multiplicity of an eigenvalue ${\lambda}$ corresponds to the number of Jordan blocks for which ${\lambda}$ is an eigenvalue.
Moreover, Jordan blocks associated with the same eigenvalue ${\lambda}$ may have different dimensions.

The ${P_{2,s}}$ are called \textit{exchange matrices}, defined as matrices with ones in the antidiagonal entries and zeros elsewhere:
\begin{align} \label{m_353}
    & P_{2,s } \coloneqq \begin{bmatrix} 
        0 & 0 & \dots & 0 & 1 \\
        0 & 0 & \dots & 1 & 0 \\
    \vdots & \vdots & \vdots & \ddots &  \vdots \\
    1 & 0 & \dots & 0 &  0 \end{bmatrix} = P_{2,s} ^{-1} = P_{2,s} ^\mathrm T ,
\end{align}
with ${\dim= s^2}$ determined by the corresponding Jordan block.

Given the matrix ${A}$ is not unique, the Hamiltonian ${H_0}$ is defined only up to a \textit{gauge transformation}---that is, a transformation that does not affect observable quantities, as we will see later when studying the time-evolution generated by ${H_0}$. The situation is, however, less clear for the interaction term ${H_1(t)}$, an issue that will remain open in this work. 

Having encoded the \ac{CPM} dynamics in terms of a Hamiltonian, we might ask for the possibility of quantizing it.
Quantization of the medium polarization sources as bosons has been proposed in Refs.~\cite{10.1021/acs.jpclett.3c01221,PhysRevLett.131.228001,sadhukhan2016quantum,jones2013quantum,Tame2013,fregoni2021strong}.
Here, imposing the \ac{CCR} 
\begin{align} \label{m_245}
    & \comm{x_\alpha}{\pi_\beta} = i \hbar \delta_{\alpha \beta},\quad \comm{x_\alpha}{x_\beta} =\comm{\pi _\alpha}{\pi_\beta}  = 0
\end{align}
means interpreting the polarization sources dynamics in \cref{m_185}, and consequently \cref{m_179}, as the Ehrenfest dynamics of the quantum operators ${x_\alpha}$, as derived in the Supplemental Material~\footnote{ See Supplemental Material (Section IV B)  at \url{http://} for a derivation of the Ehrenfest dynamics.} (see also references \cite{quintela2024quantum} therein).

\begin{figure}[h!]
\centering
    \includegraphics[width=0.4\textwidth]{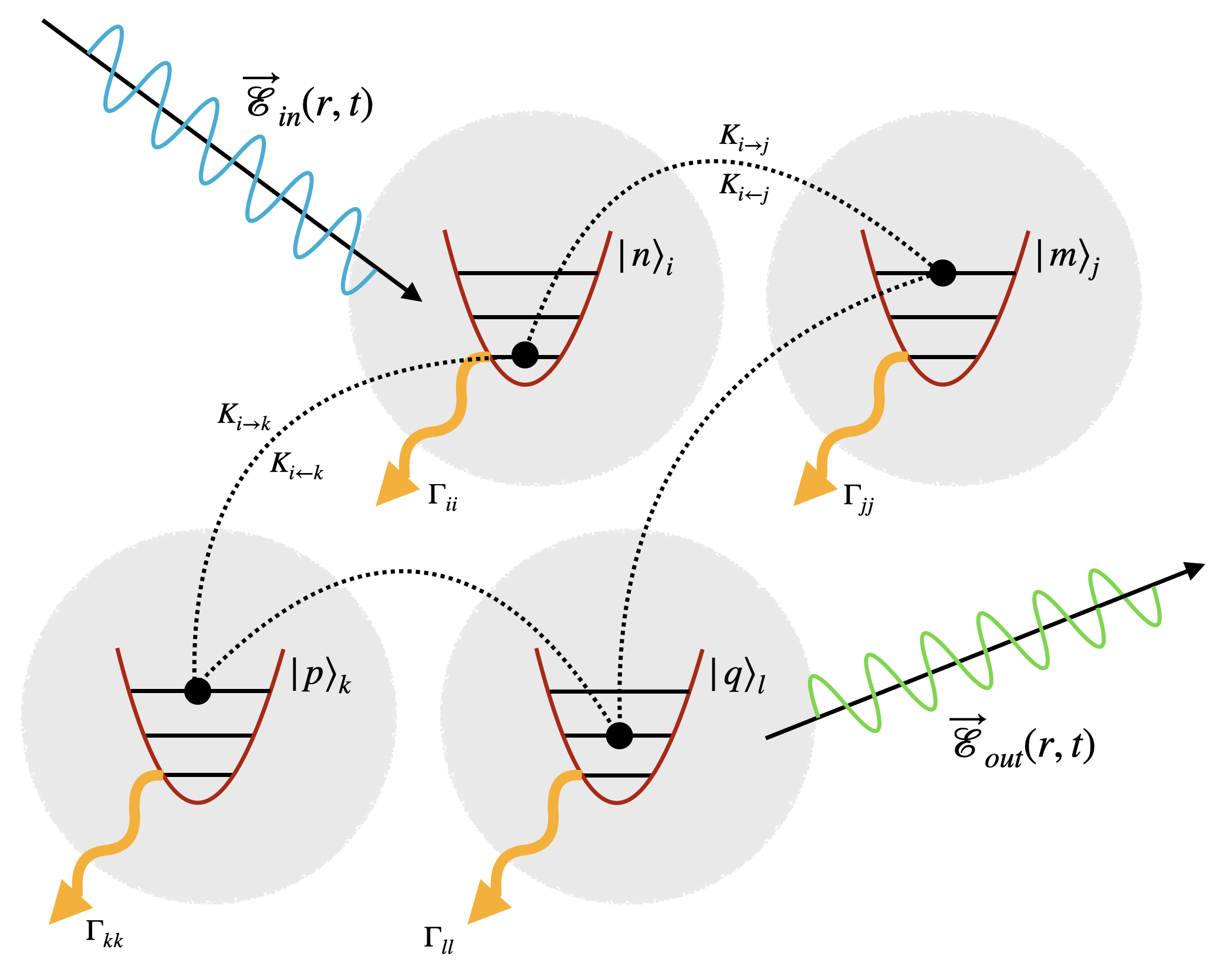}
    \caption{ \justifying Schematic representation of the \ac{QPM}. 
    It consists of a set of polarization sources whose spatial distribution is frozen (gray circles). 
    Amplitudes are quantized (represented as parabolas with discrete levels) and evolve in time according to a quadratic Hamiltonian that encodes the interaction kernel $K$ (dashed lines) and decoherence rates $\Gamma$ (wavy arrows).
    The medium interacts with a classical electric field (blue wave), polarizing the sites, which then re-emit a classical electric field (green wave).}
  \label{m_200}
\end{figure}

From now on, the media described in terms of the quantized Hamiltonian will be referred to as Quantum Polarizable Media (\ac{QPM}).
A schematic representation is provided in \cref{m_200}.
A notable property of this Hamiltonian system is that the time-independent part ${H_0}$ is not only a constant of motion but is also identically zero on-shell when solving the equations of motion for all initial conditions, both in the classical and quantum evolution.
This can be derived by combining the identity \cref{m_181} with ${\pi (t) = i A^{-1} \sqrt{\mathcal K} x(t)}$, which results from the Heisenberg equation.
In other words, the energetic contribution ${E(t) = \tr{H (t) \rho}=\tr{H_1 (t) \rho}}$ depends solely on the interaction term ${H_1(t)}$, and not on the isolated system ${H_0}$.

\subsection{Pseudo-Boson Decomposition and Bi-Coherent States} \label{m_242}

Looking to \cref{m_180}, we can write the time-independent Hamiltonian as
\begin{align} \label{m_220}
    & H_0 = \frac{1}{2} \beta^\mathrm T \beta + \frac{1}{2} \alpha^\mathrm T P_2 J_\mathcal K \alpha, 
\end{align}
where ${\beta := \sqrt{P_2} P_1 ^\mathrm T \pi}$, ${ \alpha := P_1^{-1} x}$, and 
\begin{align} 
    \label{m_224}
    & \sqrt{ P_2} = \frac{(1 + i)}{2} I + \frac{(1 - i)}{2} P_2.
\end{align}
As consequence of the \ac{CCR}, these operators verify
\begin{align} \label{m_219}
    & \comm{\beta}{\beta} = 0,\quad \comm{\alpha}{\alpha} = 0,\quad \comm{\beta}{\alpha} = -i \hbar \sqrt{ P_2}.
\end{align}
In the case of ${\mathcal K}$ diagonalizable, which is a common situation as the set of $2n$-by-$2n$ diagonalizable matrices is dense in ${\mathbb C^{2n\times 2n}}$~\cite{golub2013matrix}, results ${ J_\mathcal K = J_\mathcal K ^\mathrm T}$ and  ${P_2=I}$.
Then we can write the Hamiltonian in the canonical form 
\begin{align} \label{m_217}
    & H_0 = \hbar \tilde b^\mathrm T \sqrt {J_\mathcal K} b + \frac{\hbar}{2} \tr \sqrt {J_\mathcal K}, 
\end{align}
where ${b}$ and ${\tilde b}$ are the operators defined by
\begin{subequations} \label{m_218}
\begin{align} \label{m_229}
    & b := \frac{1 }{\sqrt{2 \hbar }} (J_\mathcal K ^{1/4}\alpha  + i J_\mathcal K ^{-1/4} \beta ) \\
    \label{m_230}
    & \tilde b :=  \frac{ 1}{\sqrt{2 \hbar }} (J_\mathcal K ^{1/4} \alpha  - i J_\mathcal K ^{-1/4} \beta ) .
\end{align}
\end{subequations}
Note that, in this representation, the spectrum of ${H_0}$ matches that of ${\sqrt{\mathcal K}}$, and we have ${\tilde b = b^\dagger }$ when ${Sp\{\sqrt \mathcal K \}\in \mathbb R}$, for example, when ${\Gamma=0}$ and ${K \geq 0}$.
This can be seen by solving $p_{\sqrt \mathcal K}(\lambda)=0$ using the formula ${\det\begin{bmatrix} A & B \\ C & D \end{bmatrix} = \det[A D - B C]}$, which is valid when ${\comm{C}{D}=0}$~\cite{Silvester_2000}.

The operators ${b}$ and ${\tilde b}$, which satisfy the commutation relations ${\comm{b_i}{\tilde b_j} = \delta_{ij}}$, are known in the literature as \textit{pseudo-boson} operators~\cite{bagarello2022pseudo}.
These can be regarded as ``ladder'' operators under the working assumption that there exist non-zero states ${\ket 0}$ and ${\ket{ 0}'}$ such that ${b \ket 0 = 0 }$ and ${(\tilde b)^\dagger \ket{ 0}' = 0 }$.
The product ${\mathrm N := \tilde b b}$, acting on the state ${\ket{ n}}$, defines a pseudo-number operator that satisfies the relations
\begin{align} \label{m_369}
    & \ket{ n} = \frac{1}{n!} (\tilde b)^n \ket{ 0}, \quad  \mathrm N \ket{ n} = n \ket{ n}.
\end{align}
However, in view of ${\tilde b \neq b^\dagger}$, these number states are not orthogonal~\cite{trifonov2009pseudobosoncoherentfockstates}.
The corresponding coherent states are defined as those satisfying
\begin{align} \label{m_368}
    & b \ket{\alpha }=\alpha \ket \alpha ,\quad (\tilde b)^\dagger \ket{ \alpha}' = \alpha \ket{ \alpha}',
\end{align}
for any ${\alpha \in \mathbb C}$~\cite{trifonov2009pseudobosoncoherentfockstates}, while the non-primed coherent state can be expanded as usual: ${\ket \alpha = e^{-\frac{1}{2}|\alpha|^2} \sum_{n=0}^\infty \frac{\alpha^n}{\sqrt{n!} }\ket n}$.

The coordinate representation of these coherent states, ${\psi_\alpha (x) = \braket{x}{\alpha}}$ and ${\varphi_\alpha (x) = \braket{x}{ \alpha }' }$, can be determined by expressing the operators ${b}$  and ${\tilde b}$ in terms of the standard coordinate operator.
Note that we are intentionally using the two symbols, ${\psi}$ and ${\varphi}$, to distinguish between the non-primed and primed states.
Their normalization factor are fixed by the bi-orthogonality condition
\begin{align} \label{m_237}
    & \braket{\psi_\alpha}{\varphi_\alpha} = 1 \Rightarrow \int_{\mathbb R^n} \dd x\, \psi^*_\alpha (x) \varphi_\alpha (x) = 1,
\end{align}
Using the above notions, we will construct the coherent states associated with ${b}$ and ${\tilde b}$, and analyze their time evolution under the Hamiltonian given in \cref{m_217}.
For further details on the theory of pseudo-boson operators, we refer the reader to Refs.~\cite{trifonov2009pseudobosoncoherentfockstates,10.1063/1.4811542,10.1063/1.3300804,10.1063/1.3514196,bagarello2022pseudo,Villegas23,Elaihar_2021,BAGARELLO2020168313,10.1063/5.0013723,Bagarello_2018}.

We can write the pseudo-boson operators ${b,\, \tilde b}$ in \cref{m_218} in terms of the coordinate and momentum operators as
\begin{subequations} \label{m_276}
\begin{align} \label{m_234}
    & b =  \frac{ 1}{\sqrt{2 \hbar }} (J_\mathcal K ^{1/4} P_1^{-1} x  + i J_\mathcal K ^{-1/4} P_1^\mathrm T \pi ) \\
    \label{m_235}
    & \tilde b =  \frac{ 1}{\sqrt{2 \hbar }} (J_\mathcal K ^{1/4} P_1^{-1} x  - i J_\mathcal K ^{-1/4} P_1^\mathrm T \pi ).
\end{align}
\end{subequations}
To find their corresponding coherent states in the coordinate representation we use \cref{m_368,m_237}, from which results
\begin{align} \label{m_236}
    & \psi_\alpha (x)= \mathcal N(\alpha) \exp{ \frac{1}{2} (x - \mu )^\mathrm T  \Sigma^{-1} (x - \mu )}\\
    \label{m_238}
    & \varphi_\alpha (x) = \mathcal N' (\alpha) \exp{- \frac{1}{2} (x - \mu )^\mathrm T  \Sigma^{-1} (x - \mu )} ,
\end{align}
where
\begin{align} \label{m_284}
    & \Sigma^{-1}:= - \hbar ^{-1} (P_1^{-1})^\mathrm T \sqrt{ J_\mathcal K} P_1^{-1}\,\text{ and } \,\mu:= P_1 J_\mathcal K^{-1/4} \alpha.
\end{align}
The normalization factors are given by 
\begin{align} \label{m_277}
    & \mathcal N^* (\alpha) \mathcal N' (\alpha) = \det{2\pi \Sigma_\text{eff}}^{-\frac{1}{2}}   \nonumber\\
    & \times \exp{\frac{1}{2}\Big [ \mu^\mathrm T \Sigma ^{-1} \mu - (\mu ^\mathrm T \Sigma ^{-1} \mu)^* - \mu_\text{eff} ^\mathrm T\Sigma_\text{eff} ^{-1} \mu_\text{eff} \Big]},
\end{align}
where ${\mu\equiv \mu (\alpha)}$ and ${\Sigma \equiv \Sigma (\alpha) }$ are functions of the amplitude of the coherent state.
The effective mean vector ${\mu_\text{eff}}$ and covariance matrix ${\Sigma_\text{eff}}$ are defined as follows:
\begin{subequations} \label{m_282}
\begin{align} \label{m_279}
    & \Sigma_\text{eff} := \big( - (\Sigma^{-1})^* + \Sigma^{-1} \big)^{-1} \\
    \label{m_280}
    & \mu_\text{eff} :=\Sigma_\text{eff} (- (\Sigma^{-1})^*\mu^*  + \Sigma^{-1} \mu ).
\end{align}
\end{subequations}
An easily verifiable fact is that evolution under the Hamiltonian in \cref{m_217} preserves the coherent character of the state~\cite{trifonov2009pseudobosoncoherentfockstates}.
Consequently, the evolved state can be determined by substituting ${\alpha \mapsto e^{-i t \sqrt{ J_\mathcal K}} \alpha}$ into \cref{m_236,m_238,m_277,m_282}, along with a global multiplication by the exponential factor ${\exp{-\frac{i t}{2} \tr \sqrt {J_\mathcal K} - \frac{1}{2}\Big( |\alpha|^2 - |e^{-i t \sqrt {J_\mathcal K}} \alpha|^2 \Big)}}$, namely
\begin{align} \label{m_283}
    & \psi_\alpha (x) \mapsto e^{-\frac{i t}{2} \tr \sqrt {J_\mathcal K} - \frac{1}{2}\Big( |\alpha|^2 - |e^{-i t \sqrt {J_\mathcal K}} \alpha|^2 \Big)} \psi_{e^{-i t \sqrt{ J_\mathcal K}} \alpha} (x).
\end{align}
The complex amplitude of ${\sqrt{ J_\mathcal K}}$ induces two kinds of damping (given ${\Im \sqrt {J_\mathcal K } < 0}$) of the unnormalized coherent state: 
1. Damping of ${\psi_\alpha(x)}$ into the ground state ${\psi_0 (x)}$ (i.e., the coherent state with ${\alpha=0}$); and 
2. Global damping of the overall amplitude to zero.
This latter effect is absorbed by the normalization coefficients ${\mathcal N (\alpha _t)}$ and ${\mathcal N (\alpha _t)'}$.
In pseudo-Hermitian mechanics, the time evolution of the non-primed state ${\psi_\alpha}$ and primed state ${\varphi_\alpha}$ is governed by the operators ${e^{-i/\hbar H_0 t}}$ and ${e^{-i/\hbar H_0 ^\dagger t}}$, respectively~\cite{trifonov2009pseudobosoncoherentfockstates}. 
Consequently, the normalization coefficients grow exponentially with increasing time, specifically ${\mathcal N^*(\alpha _t) \mathcal N'(\alpha _t) \propto \exp{- t \tr [\Im \sqrt {J_\mathcal K }]} }$.

\section{Applications} \label{m_244}

\subsection{Polarization Signal from the Eigenvalues and Eigenvectors of ${\sqrt \mathcal K}$} \label{m_327}

The quantity commonly used to measure the spectral information of polarizable sources is the imaginary part of the polarizability function, defined as:
\begin{align} \label{m_248}
    & \Im \alpha_{\vec s} (\omega) := \Im { \sum_{\beta=1}^n R_{s_\beta \beta}^{(u)} u _{\beta} (\omega) },
\end{align}
where ${R_{\vec s}^{(u)} := \begin{bmatrix} R_{s_1 1}^{(u)}, \dots,\, R_{s_\beta \beta }^{(u)}, \dots,\, R_{s_n n}^{(u)} \end{bmatrix}^\mathrm T }$ is the generalized coordinate vector and ${u_\alpha(\omega)}$ are the polarization sources. By using the superscript ${(u)}$, we are emphasising the dependence of the quantity ${R_{\vec s}}$ on the type of polarizable sources. Whenever the meaning is clear from the context, we will omit it to simplify the notation.

The derivation of the optical response for a general class of polarizable sources---including charges, dipoles and multipoles as dynamical variables---requires proper specification of the quantities appearing in \cref{m_248}.
For example, when ${u_\alpha (\omega)}$ represent charges, then ${R_{\vec s}^{(u)} = R_{\vec s} }$ correspond to fixed spatial coordinates, where ${s_\beta \in \{1,2,3\}}$ indicates the spatial component of the coordinate vector.
In other cases, the generalized coordinate vector list other quantities which, when multiplied by the specific type of polarizable source, ensure the correct dimensionality in \cref{m_248}.
In the $\omega$FQF$\mu$ model, for instance, we have~\footnote{See the definition of the complex polarizability in the Supplemental Material of Ref.~\cite{giovannini2022we}, Eqs.~(S10) and (S11).}
\begin{align} \label{m_371}
    & R_{\vec s}^{(u)} = \begin{cases} R_{\vec s}\, & u_\alpha \text{ is a charge} \\ 1_{\vec s}\, & u_\alpha \text{ is a dipole} \end{cases}.
\end{align}
The polarizable sources are expressed in the frequency domain as:
\begin{align} \label{m_295}
    & u_\beta (\omega ) = \sum_{\nu =1}^n A_{\beta \nu } (\omega) f_\nu (\omega) \\
    \label{m_294}
    & A(\omega):= ( \omega^2 + 2 i \omega \Gamma - K ) ^{-1} \in \mathbb C ^{n\times n},
\end{align}
where ${f(\omega) = \begin{bmatrix} f_1 (\omega), \dots,\, f_{\beta }(\omega), \dots,\, f_{n}(\omega) \end{bmatrix}^\mathrm T \in \mathbb C^{n\times 1}}$.

We can rewrite \cref{m_248} in a more compact way using the tensor product and trace operations,
\begin{align} \label{m_326}
    & \Im[ \alpha_{\vec s} (\omega)] = \Im[ \tr{A(\omega) f(\omega) \otimes R_{\vec s}} ].
\end{align}
In \cref{m_356}, we have shown that all the spectral information of ${A(\omega)}$ is given by ${\sqrt \mathcal K \in \mathbb C^{2n \times 2n}}$.
Within our newly developed formalism, we can compute the polarizability function as follows:
\begin{align} \label{m_308}
    & \Im \alpha_{\vec s} (\omega) \nonumber\\
    & = \Im\Bigg[ \tr{ (\omega^2 - \mathcal K)^{-1} \begin{bmatrix} I & 0 \\ 0 & - i \omega - 2 \Gamma \end{bmatrix} 
        \begin{bmatrix} f(\omega) \\ f(\omega) \end{bmatrix} \otimes  \begin{bmatrix} R_{\vec s} \\ 0 \end{bmatrix}}\Bigg] .
\end{align}
From the Jordan decomposition ${\mathcal K} = P J P^{-1}$ and using the cyclic property of the trace we have 
\begin{align} \label{m_303}
    & \Im \alpha_{\vec s} (\omega) = \Im \tr{ (\omega^2 - J)^{-1} C_{\vec s} (\omega)  } \\
    & C_{\vec s} (\omega) :=P^{-1} \begin{bmatrix} f(\omega) \otimes R_{\vec s} & 0 \\ - ( i \omega + 2 \Gamma) (f(\omega) \otimes R_{\vec s}) & 0 \end{bmatrix} P.
\end{align}
From now on we will assume that ${J}$ is diagonal. 
To account for the ${\Im[\cdot]}$ function in the above expressions, it is convenient to introduce the following definitions
\begin{subequations} \label{m_311}
\begin{align} \label{m_304}
    & (\omega^2 - J)^{-1}_k = \mathcal D_k (\omega) + i \mathcal A_k (\omega) \\
    \label{m_306}
    & \mathcal D_k (\omega):= \frac{\omega^2 - \lambda_k'}{(\omega^2 - \lambda_k' )^2 + (\lambda_k'' )^2} \\
    \label{m_305}
    & \mathcal A_k (\omega): = \frac{\lambda_k''}{(\omega^2 - \lambda_k' )^2 + (\lambda_k'' )^2},
\end{align}
\end{subequations}
where ${\mathcal A_k (\omega)}$ and ${\mathcal D_k (\omega)}$ stand for the absorptive and dispersive contributions of the ${k}$th-eigenvalue of ${J}$, namely ${\lambda_k = \lambda_k' + i \lambda_k '' }$.
Using the identities ${\Im[ab] = \Im[a] \Re[b] + \Re[a]\Im[b] }$ and ${\Im \big[\sum_k a_k \big] = \sum_k \Im[a_k]}$ we have 
\begin{align} \label{m_307}
    & \Im \alpha_{\vec s} (\omega) = \sum_{k=1}^{2n} \mathcal A_k (\omega) S_{\vec{s} k}^\mathcal A (\omega) + \mathcal D_k (\omega) S_{\vec{s} k}^\mathcal D (\omega),
\end{align}
where
\begin{subequations} \label{m_313}
\begin{align} \label{m_309}
    & S_{\vec{s} k}^\mathcal A (\omega): = \Re[C_{{\vec s},kk} (\omega)] \\
    \label{m_314}
    &  S_{\vec{s} k}^\mathcal D (\omega): = \Im[C_{{\vec s},kk} (\omega)]\\
    \label{m_312}
    & \sum_{k=1}^{2n} S_{\vec{s} k}^{\mathcal A } (\omega) + S_{\vec{s} k}^{ \mathcal D} (\omega) = \tr{f(\omega) R_{\vec{s}} ^\mathrm T} .
\end{align}
\end{subequations} 
Note that in \cref{m_307}, the coefficients ${S_{\vec{s} k}^\mathcal A (\omega)}$ and ${S_{\vec{s} k}^\mathcal D (\omega)}$ represent the amplitudes of the absorptive and dispersive contributions of ${\lambda_k}$, respectively. 
These coefficients exhibit two meaningful properties:
\begin{enumerate}
    \item they depend on all the eigenvectors ${P}$ of ${\mathcal K}$, which is indicative of a collective phenomenon, but are independent of the eigenvalues;
    \item they are in a one-to-one correspondence to the eigenvalues, allowing us to use them to ``weight'' the contributions of ${\lambda_k}$ to the polarization.
\end{enumerate}
If the external force acting on the polarization sources is a kick pulse, i.e. $f(\omega) = f$ is a frequency-independent constant, the coefficients ${S_{\vec{s} k}^\mathcal A (\omega)}$ and ${S_{\vec{s} k}^\mathcal D (\omega)}$ become a linear function of the frequency $\omega$, expressed as follows:
\begin{subequations} \label{eq:decomposition}
\begin{align} \label{eq:dec1}
    S_{\vec{s} k}^\mathcal A (\omega) & = \Re[I_{k}] - i\omega \Im[A_{k}]\\
    \label{eq:dec2}
    S_{\vec{s} k}^\mathcal D (\omega) & = \Im[I_{k}] - i\omega \Re[A_{k}]
\end{align}
\end{subequations} 
where ${I, A \in\mathbb{C}^{2n}}$ are the intercept and the angle of the linear function, defined by:
\begin{subequations} \label{m_357}
\begin{align} 
    & I_k := \begin{bmatrix} I_{k,1} & I_{k,2} \end{bmatrix}^\mathrm T  \\
    & I_{k,1} := [\big( (P^{-1})_{11} - 2 (P^{-1})_{12} \Gamma\big) (f \otimes R_{\vec s}) P_{11}]_{kk} \label{}\\
    & I_{k,2} : = [\big( (P^{-1})_{21} - 2 (P^{-1})_{22} \Gamma\big) (f \otimes R_{\vec s}) P_{12}]_{kk} \label{}
\end{align}
\end{subequations} 
and 
\begin{subequations} \label{m_358}
\begin{align} 
    & A_k := \begin{bmatrix} A_{k,1} & A_{k,2} \end{bmatrix}^\mathrm T \\
    & A_{k,1} : = [ (P^{-1})_{12}\, (f \otimes R_{\vec s}) P_{11}]_{kk} \label{}\\ 
    & A_{k,2} : = [ (P^{-1}) _{22}\, (f \otimes R_{\vec s}) P_{12}]_{kk} \label{},
\end{align}
\end{subequations}
where ${P = \begin{bmatrix} P_{11} & P_{12} \\ P_{21} & P_{22} \end{bmatrix}}$.
For a derivation of these equations see the Supplemental Material~\footnote{See Supplemental Material (Section V) at \url{http://} for the derivation of \cref{eq:decomposition}.}.

\subsection{Polarization Signal of Plasmonic Nanostructures} \label{m_328}

\begin{figure*}[h!] 
     \centering
     \begin{subfigure}[t]{0.23\textwidth}
         \centering
         \includegraphics[width=\textwidth]{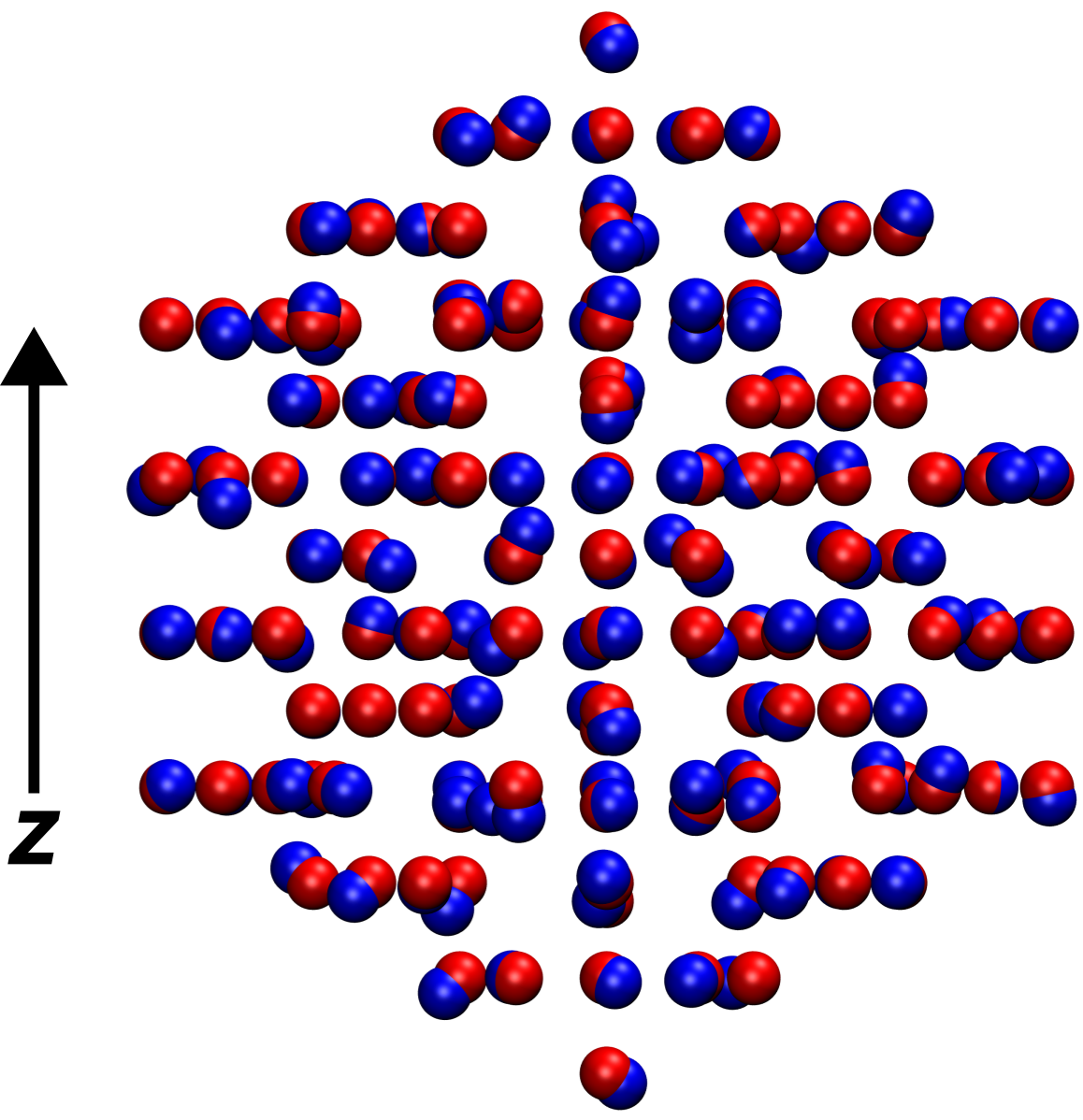}
         \caption{Ag147}
    \label{m_329}
     \end{subfigure}
     \hfill
     \begin{subfigure}[t]{0.3\textwidth}
         \centering
         \includegraphics[width=\textwidth]{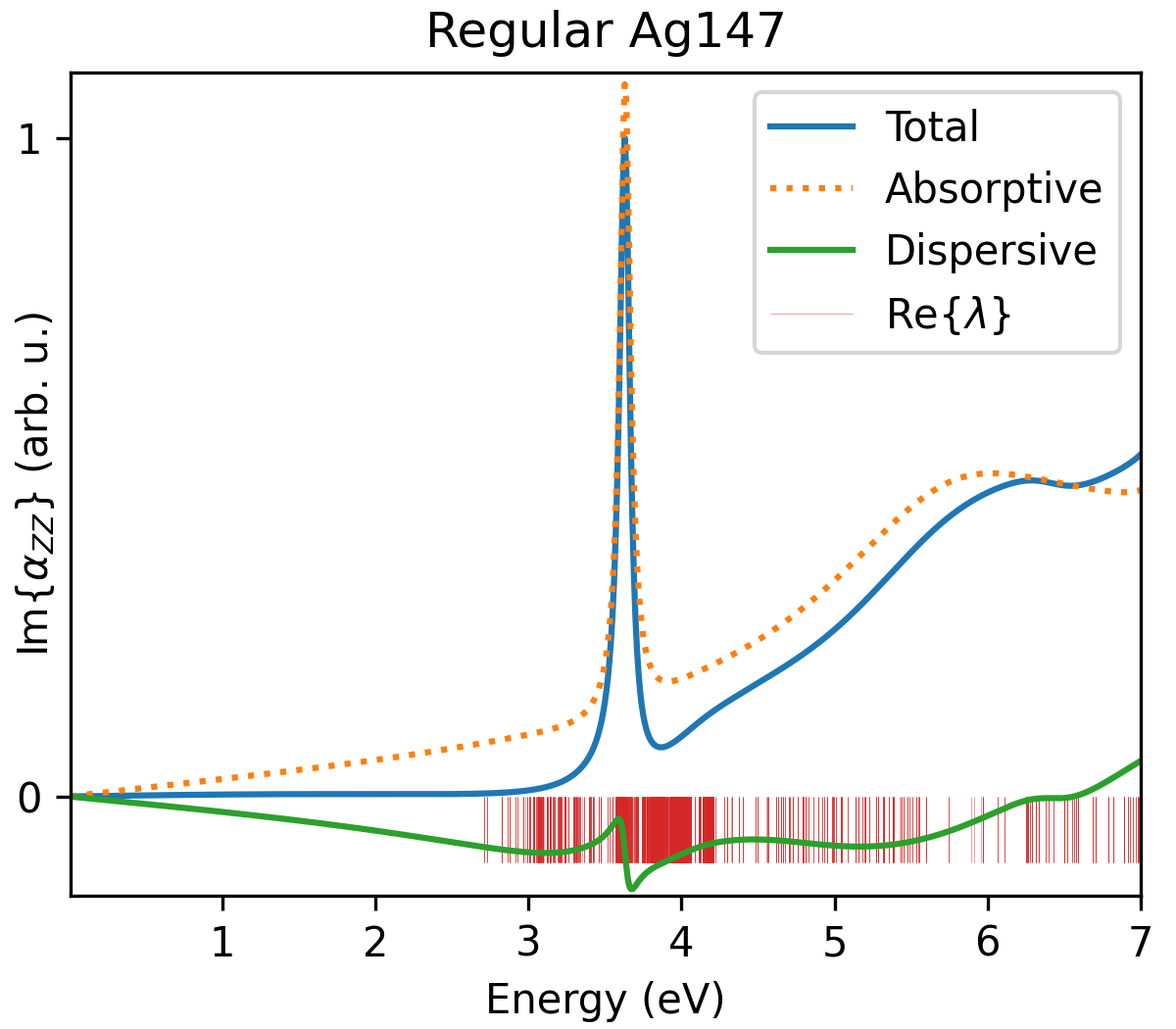}
         \caption{\justifying }
    \label{m_330}
     \end{subfigure}
     \hfill
     \begin{subfigure}[t]{0.3\textwidth}
         \centering
         \includegraphics[width=\textwidth]{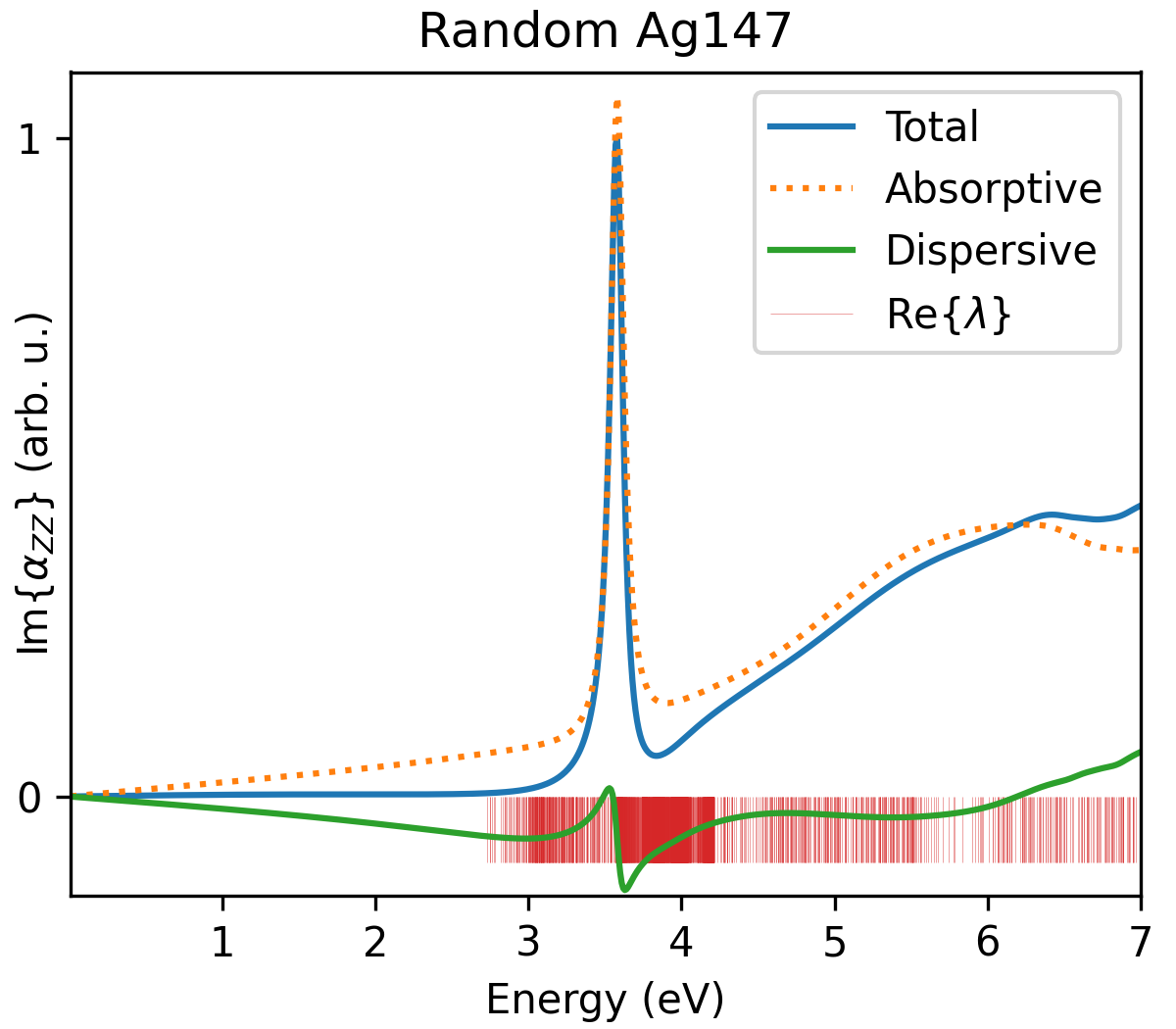}
         \caption{\justifying }
    \label{m_331}
     \end{subfigure}
     
     \begin{subfigure}[t]{0.23\textwidth}
         \centering
         \includegraphics[width=\textwidth]{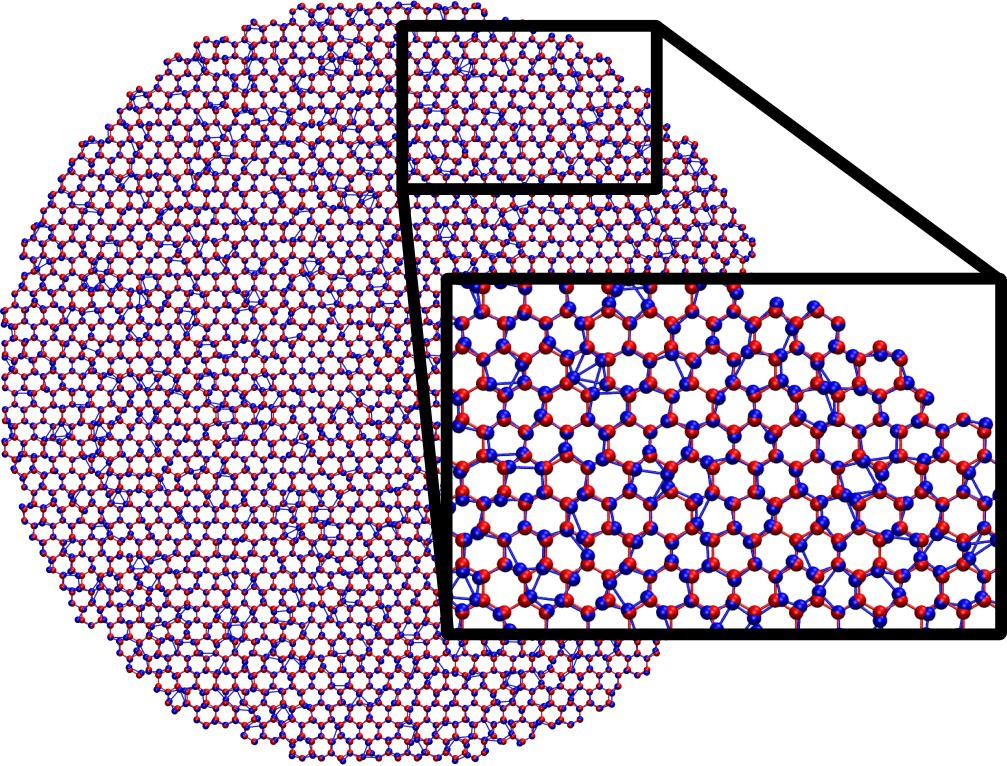}
         \caption{GD10}
    \label{m_332}
     \end{subfigure}
     \hfill
     \begin{subfigure}[t]{0.3\textwidth}
         \centering
         \includegraphics[width=\textwidth]{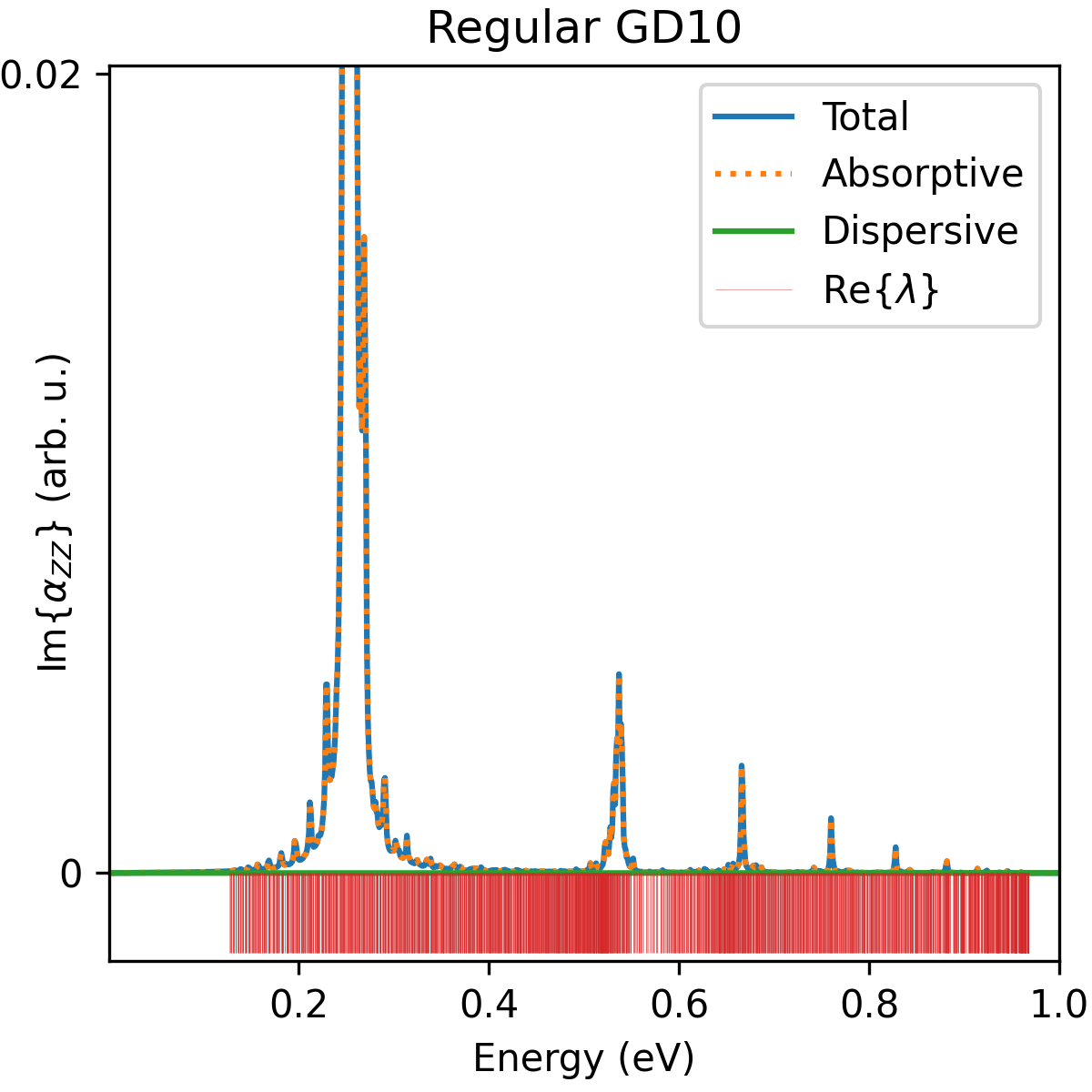}
         \caption{\justifying }
         \label{m_333}
     \end{subfigure}
     \hfill
     \begin{subfigure}[t]{0.3\textwidth}
         \centering
         \includegraphics[width=\textwidth]{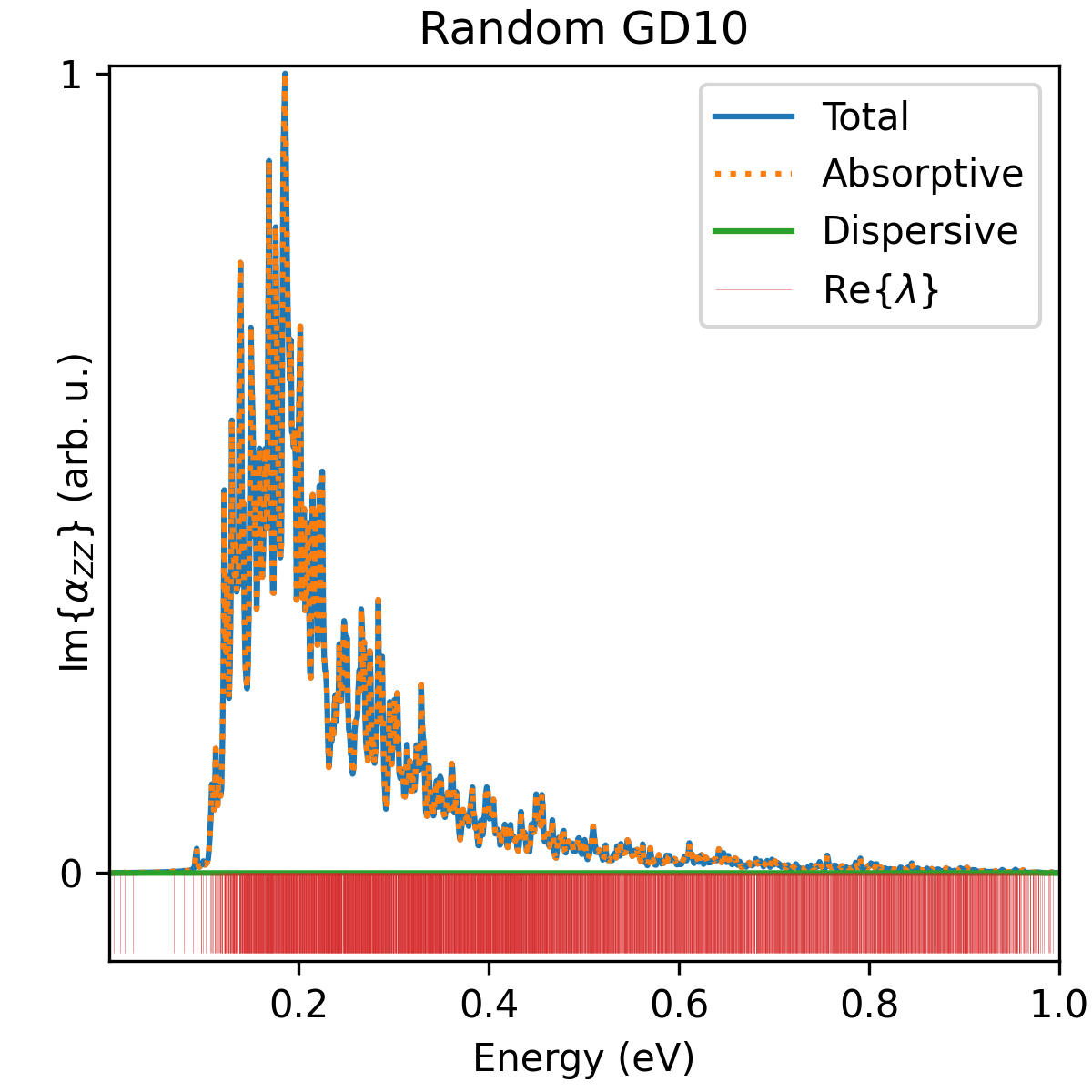}
         \caption{\justifying }
    \label{m_334}
     \end{subfigure}
         \caption{\justifying 
    Simulated optical response of two plasmonic systems: a Silver cluster composed of 147 atoms in an icosahedral shape (\ce{Ag147}; \cref{m_329}) and a graphene disk with a diameter of 10 nm (GD10; \cref{m_332}).
    Red and blue atoms indicate the standard and perturbed structures, respectively, with the latter incorporating noise in the atomic positions.
The simulated signal consists of the $zz$-component of the imaginary part of the polarizability function, plotted along the vertical axis as a function of the incident field frequency, shown on the horizontal axis.
    The optical response of \ce{Ag147} is simulated in \cref{m_330,m_331} using the $\omega$FQF$\mu$ method, following the  parametrization  proposed in Refs.~\cite{giovannini2022we,lafiosca2024real}.
    Similarly, in \cref{m_333,m_334}, the optical response of GD10 is simulated using the $\omega$FQ method, with its parametrization reported in Ref.~\cite{giovannini2020graphene}.
    The plotted signals differentiate between the Absorptive and Dispersive contributions, while the real part of the eigenvalues of ${\sqrt \mathcal K}$ is shown as red vertical lines at the bottom of each figure.}
        \label{m_335}
\end{figure*}

In this section, we study the optical response of two plasmonic systems: a Silver cluster composed of 147 atoms in an icosahedral shape (\ce{Ag147}) and a graphene disk with a diameter of 10 nm (GD10), depicted in \cref{m_329,m_332}, respectively. 
For each system, we consider a ``regular'' structure with the highest symmetry and a ``random'' structure, obtained by perturbing each atom's position with a random vector in the space for \ce{Ag147} and on the graphene plane for GD10, with a maximum displacement of 0.5 \AA. 
In \cref{m_329,m_332}, the regular structure is represented by red atoms, while blue atoms represent the random structure.
The optical response of \ce{Ag147} is simulated using the $\omega$FQF$\mu$ model with parameters proposed in Refs.~\cite{giovannini2022we,lafiosca2024real}, while for the GD10 we use the $\omega$FQ method with parameters proposed in Ref.~\cite{giovannini2020graphene}.
GD10 contains 2990 Carbon atoms, each represented in the $\omega$FQ approach by a single charge, leading to a matrix order of 5980. In the case of \ce{Ag147}, in the real-time formulation of the $\omega$FQF$\mu$ method, each Silver atom is described by one charge and five dipoles, which represent the partitioning of the interband polarizability in terms of Drude-Lorentz oscillators~\cite{lafiosca2024real}.
Therefore, we have 16 coordinates for each atom, leading to a total matrix order of $4704$. 

We calculate the $zz$ component of the imaginary part of the polarizability function, where the $z$ axis is indicated in \cref{m_329}. 
The function ${\Im \alpha_{zz}(\omega)}$ is obtained from \cref{m_248} by fixing the direction of the polarizable sources along the ${z}$-axis, i.e., the third component ${z\leftrightarrow 3}$, so that ${(\vec s)_{k \beta} = \delta_{k 3} s_\beta }$.
The result is then divided by the amplitude of the incident field ${E_{0,z}}$, which is also oriented along the ${z}$-axis. That is,
\begin{align} \label{m_372}
    &  \Im \alpha_{zz}^{\omega FQ} (\omega) \coloneqq \Im \Bigg\{ \sum_{\beta=1}^n R^{(u)}_{3 \beta} u_\beta (\omega)\Bigg\} /E_{0,z}.
\end{align}
For ${\omega}$FQ model, the polarizable sources ${u_\beta (\omega)}$ represent the charges ${c_\beta (\omega)}$, and the generalized coordinate vector components ${R^{(u)}_{3 \beta} }$ reduce to the spatial positions ${R_{3 \beta}}$.
For the ${\omega}$FQF${\mu}$ model, the number of polarizable sources is doubled, as both the charges ${c_\beta (\omega)}$ and dipoles ${\mu_\beta (\omega)}$ are included. The imaginary part of the polarizability function in this case reads:
\begin{align} \label{m_373}
    & \Im \alpha_{zz}^{\omega FQF \mu} (\omega) \coloneqq  \Im \Bigg\{ \sum_{\beta=1}^n  R_{3 \beta} c_\beta (\omega) + \sum_{\beta=1}^n \mu_\beta (\omega) \Bigg\} /E_{0,z}.
\end{align}

The $z$-axis serves as a symmetry axis for \ce{Ag147} and lies on the molecular plane for GD10.
The investigated frequency range is from 0 to 7 eV with a step of 0.01 eV for \ce{Ag147} and from 0 to 1 eV with a step of 0.001 eV for GD10. 
In \cref{m_330,m_333} we report $\Im \alpha_{zz} (\omega)$ normalized to $1$ in the given frequency range, as calculated for the regular \ce{Ag147} (solid blue lines). 
The real part of the eigenvalues of $\sqrt{\mathcal{K}}$ is reported with red vertical lines at the bottom of each figure.
We notice that the polarizability function is characterized by a strong plasmon resonance absorption peak at 3.63 eV~\cite{giovannini2022we}.
In comparison, from 4 eV and above the interband absorption mechanism dominates the spectral response leading to a broad and increasing absorption band~\cite{giovannini2022we}.
The real part of the eigenvalue distribution of $\sqrt{\mathcal{K}}$ is mostly concentrated close to the plasmon peak. 

From \cref{m_307}, we can investigate what is the relative contribution of the absorptive and dispersive portions, i.e. in which only the $\mathcal A_k (\omega) S_{\vec{s} k}^\mathcal A (\omega)$ or $\mathcal D_k (\omega) S_{\vec{s} k}^\mathcal D (\omega)$ terms are retained in the summation over the eigenpairs, respectively. The absorptive contribution is reported in\cref{m_330} with a dotted orange line, while the solid green line represents the dispersive contribution. 
It can be noticed that for the regular \ce{Ag147} the absorptive part dominates the polarization function and it is always positive on the frequency range, while the dispersive part introduces a relatively small and negative correction, with an inflection point on the plasmon peak.
In the case of GD10, the polarizability function in \cref{m_333} presents a strong absorption peak at 0.255 eV, and a series of little features at higher frequencies, which can be associated with high-order multipolar plasmon resonances~\cite{lafiosca2021going}. 
We notice that the eigenvalue distribution of $\sqrt{\mathcal K}$ is dense and seemingly uniform at given regions, and that the dispersive contribution to the polarization function is negligible across the whole frequency range. 

Moving to the random case, we notice that for \ce{Ag147} in \cref{m_331} the geometric perturbation has a limited impact on the polarizability function, in particular by shifting the plasmon absorption peak at 3.58 eV and by slightly increasing the relative contribution of the dispersive portion.
Furthermore, the eigenvalue distribution becomes more uniform in the frequency range, especially between 3 and 4 eV.
On the other hand, the randomization of the GD10 atoms in the molecular plane has a huge effect on the polarizability signal as reported in \cref{m_334}.
Here, we observe a broad and crowded absorption band with a maximum located at 0.186 eV.  \\
As the next step, we investigate the possibility of reproducing the polarization signal of the chosen plasmonic nanostructures using reduced dimensionality, i.e., by selecting a subset of the eigenvectors and eigenvalues in \cref{m_307}.
To achieve this, we consider two filtering approaches for selecting specific eigenpairs:
1) \ac{EF}, in which the pair $(\lambda_k, v_k)$ is selected if $|\Re \lambda_k|$ lies in a given frequency range; and
2) \ac{IF}, in which the pair $(\lambda_k, v_k)$ is selected if $|\Re I_k|$ exceeds a given threshold, with ${I_k}$ given by \cref{m_357}. 
The \ac{EF} is in principle able to select all the relevant contributions of the spectrum to the polarization function in a given frequency range, given the decay properties of ${\mathcal D_k (\omega)}$ and ${\mathcal A_k (\omega)}$ (see \cref{m_306,m_305}). 
On the other hand, we have shown that in the case of a kick pulse, the absorptive contribution dominates the polarization function over the whole frequency range for \ce{Ag147} and GD10, therefore ${S_{\vec{s} k}^\mathcal A (\omega)}$ and thus $\Re\{I_k\}$ can be used to select only the most relevant eigenvectors (see \cref{eq:dec1,eq:dec2}). \\
\begin{figure}[h!]
\centering
    \includegraphics[width=0.45\textwidth]{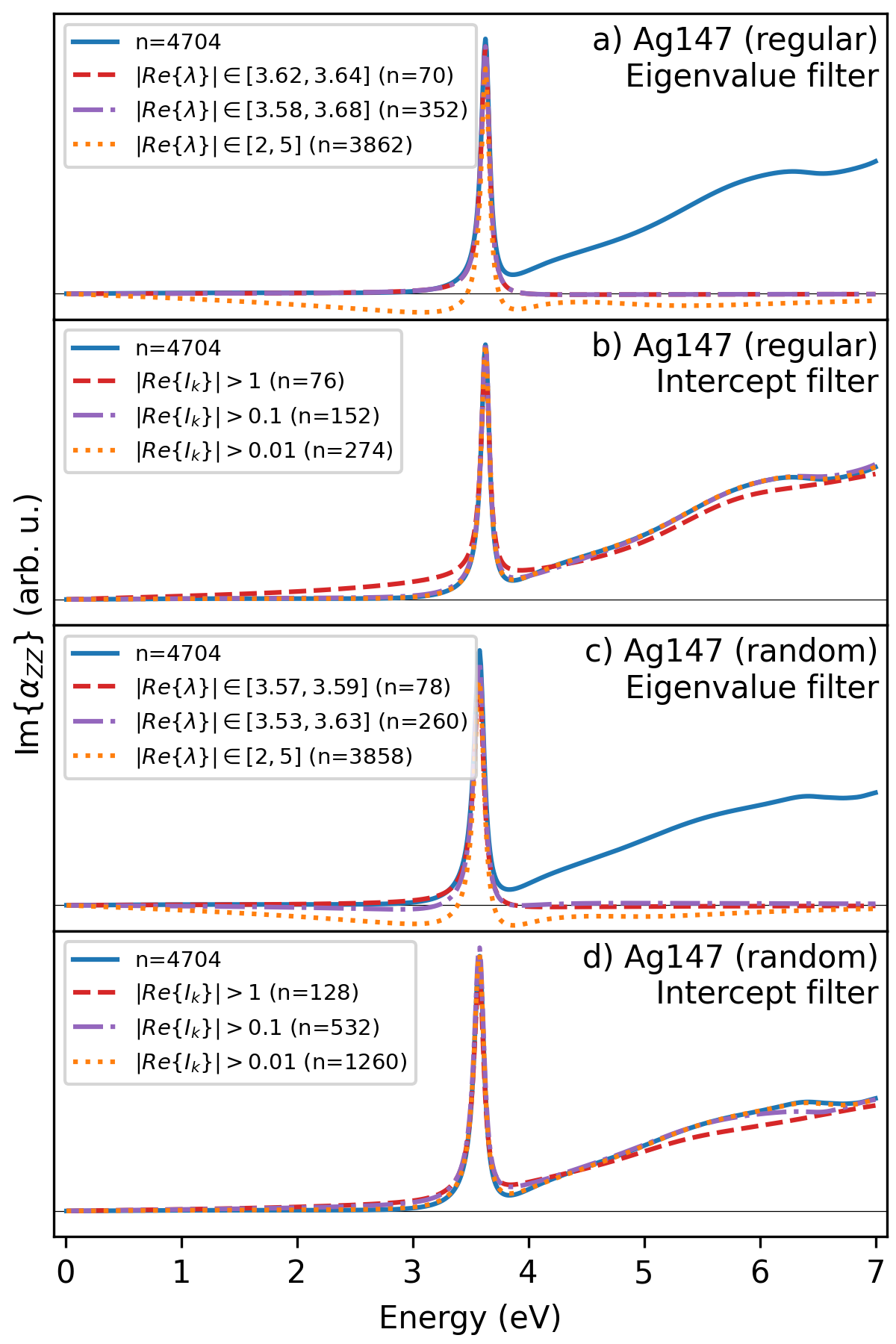}
    \caption{ \justifying Reconstruction of the optical response of the icosahedral \ce{Ag147} structure by filtering mode contributions to the signal, using various frequency ranges for the \ac{EF} (panels a,c) and thresholds for \ac{IF} (panels b, d).
    The signals of the standard and perturbed structure are shown in panels (a,b) and panels (c,d), respectively.}
    \label{m_345}
\end{figure}
In \cref{m_345} we report the reconstruction of the polarizability function by using various frequency ranges for the \ac{EF} (panels a,c) and thresholds for \ac{IF} (panels b, d) in the case of \ce{Ag147} structures. 
We start with the \ac{EF} on the regular \ce{Ag147} structure, reported in panel a of \cref{m_345}. 
The frequency range is centered on the plasmon absorption peak at 3.63 eV, and we notice that 70 eigenvalues (dotted red line) are sufficient to reconstruct the absorption peak also from the quantitative point of view, and no improvement is obtained by taking 352 eigenvalues (dashed purple line). Outside the filtering frequency range the polarization function becomes rapidly zero, and no contributions to the interband portion are obtained. 
Moreover, we notice that the \ac{EF} does not converge uniformly with the number of eigenvalues, as enlarging the filter range from 2 to 5 eV (dotted orange line) causes the polarizability function to acquire some negative contributions around the plasmon peak, worsening the agreement with the reference function.\\
Moving to the \ac{IF} approach, in panel b of \cref{m_345} we notice that 76 eigenvalues (dashed red line) are sufficient to have a good description of the absorption profile across the whole frequency range since we are selecting the largest values of $S_{\vec{s} k}^\mathcal A (\omega)$ contributing to the polarization function. Moreover, the \ac{IF} approach can be systematically improved by reducing the threshold value, and a uniform convergence is achieved.\\
In the case of a perturbed \ce{Ag147} structure, for the \ac{EF} approach (panel c of \cref{m_345}), similar considerations to those of the regular \ce{Ag147} structure apply.
Here, 78 eigenvalues concentrated near the plasmon peak at 3.58 eV are sufficient to achieve a quantitative match of the polarizability function in the immediate surroundings, although the convergence remains non-uniform as the number of eigenvalues increases.
Moving to the \ac{IF} approach (panel d of \cref{m_345}), we observe that using the same thresholds as for the regular \ce{Ag147} results in more eigenvalues being extracted.  In the smallest batch, 128 eigenvalues are obtained (dashed red line).
This can be attributed to the symmetry reduction in the perturbed \ce{Ag147} structure compared to the regular \ce{Ag147} structure.
\begin{figure}[h!]
\centering
    \includegraphics[width=0.45\textwidth]{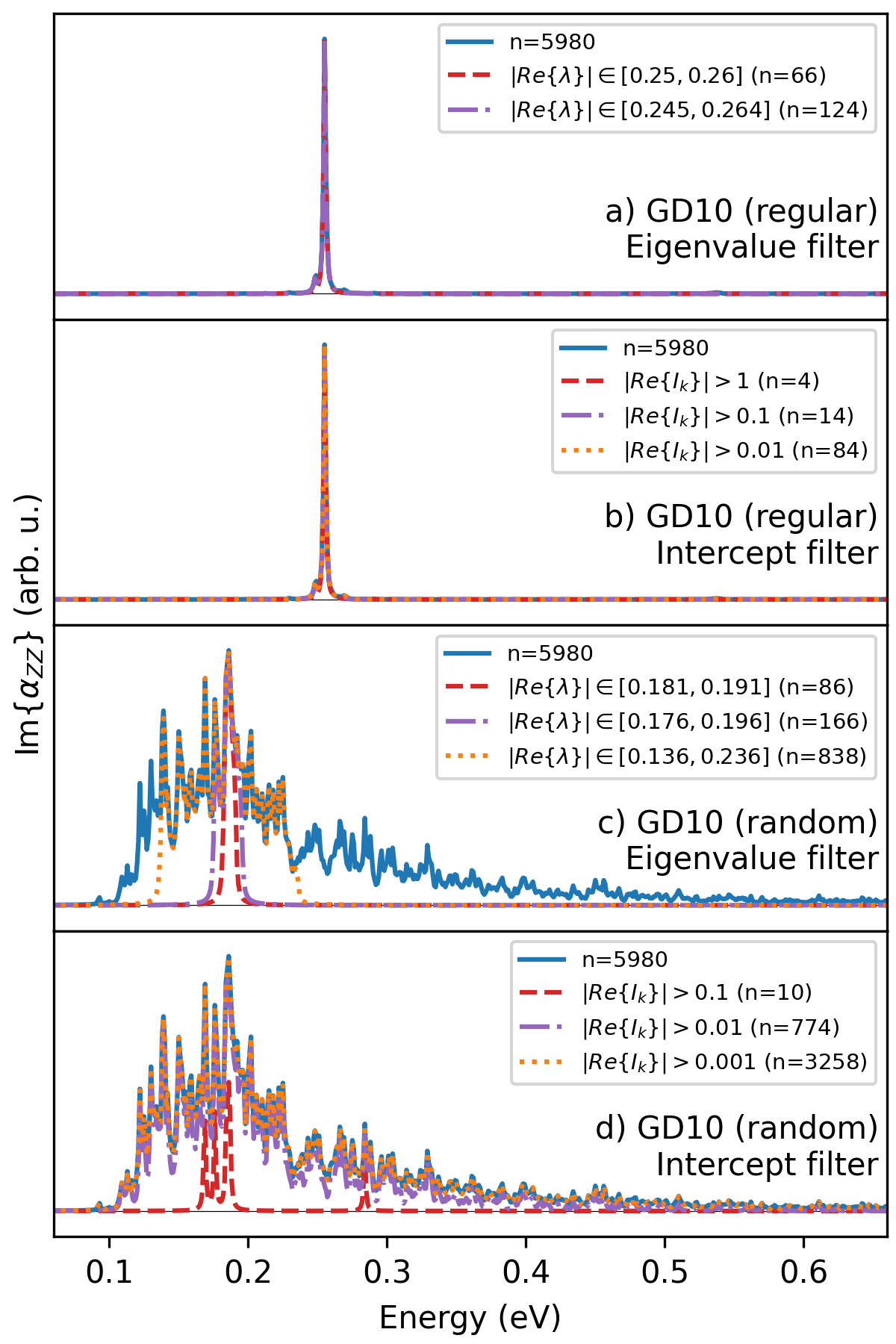}
    \caption{ \justifying Reconstruction of the optical response of the GD10 structure by filtering mode contributions to the signal, using various frequency ranges for the \ac{EF} (panels a,c) and thresholds for \ac{IF} (panels b, d).
    The signals of the standard and perturbed structure are shown in panels (a,b) and panels (c,d), respectively.}
    \label{m_345_bis}
\end{figure}
In \cref{m_345_bis} we report the reconstruction of the polarizability function for the GD10 structures. 
Starting with the regular GD10 system (panels a, b), we observe that both \ac{EF} and \ac{IF} approaches can be effectively adopted to quantitatively reproduce the polarization function at the absorption peak located at 0.255 eV, requiring only a limited number of eigenvalues--- up to just 4 by using the \ac{IF} method. \\
In the case of the perturbed GD10 structure, the \ac{EF} approach (panel c) adopted using larger frequency ranges leads to a better description of the absorption band locally, and the polarization function quickly becomes zero outside the selected range. 
Here, the \ac{EF} approach can be systematically improved, and a monotonic convergence is observed, in contrast to the \ce{Ag147} case. 
On the other hand, the \ac{IF} approach (panel d) allows for a more uniform description of the polarizability function across the whole frequency range, even though more than 10 \% of the eigenpairs are needed to have a good description, as with a threshold value of 0.01 (dashed purple line, 774 eigenvalues). \\

\section{Self-consistent electrical field in the semiclassical approximation} \label{m_273}
\subsection{\ac{QPM} Dynamics in Phase Space} \label{m_247}

The quantum dynamics of the \ac{QPM} over arbitrary \ac{GS} can be effectively analyzed using the phase-space representation, which simplifies the description of the dynamics. 
In the Schrödinger picture, a time-dependent Hamiltonian ${H(t)}$ governs the evolution as follows:
\begin{align} \label{m_172}
    & \rho _t =  U _t \rho_0  U^{-1} _t \\
    \label{m_173}
    & U _t = \mathcal T e^{ -i /\hbar \int _0 ^t  H (\tau ) \dd \tau } \\
    & \expval{ \mathcal O}_t = \tr {  \mathcal O \rho _t }, 
\end{align}
where ${\rho_t}$ is the density matrix with the time dependence is indicated by the subscript $t$, ${\mathcal O}$ is an arbitrary operator and ${U_t}$ the evolution operator with ${ \mathcal T}$ indicating the synchronously time-ordered exponential~\cite{mukamel1995principles}.
The replacement of ${U_t ^\dagger}$ with ${U_t ^{-1}}$ is motivated by the possibility of non-Hermitian Hamiltonian evolutions~\cite{akhundova1982wigner}. 
Moreover, it can be verified that the density matrix in \cref{m_172} evolves according to the von Neumann equation ${\dv{}{t}\rho(t) = - \frac{i}{\hbar} \comm{H(t)}{\rho (t)}}$~\footnote{See Supplemental Material (Section VI) at \url{http://} for a derivation of the von Neumann equation for non-Hermitian Hamiltonians.}.
Defining the canonical operator as ${q := \begin{bmatrix} \pi & x \end{bmatrix}^\mathrm T }$ with ${\dim = 2N}$, where ${N=2n}$, and the matrices 
\begin{align} \label{m_182}
    & \mathcal B  :=  \begin{bmatrix} A & 0 \\ 0 & A^{-1} \mathcal K \end{bmatrix} \\
    \label{m_183}
    & \mathcal{C}_t := \begin{bmatrix} 0 & A^{-1} F_t \end{bmatrix}^\mathrm{T} , 
\end{align}
we can rewrite the \ac{QPM} Hamiltonian (\cref{m_180}) as 
\begin{align} \label{m_184}
    & H(t) = \frac{1}{2} q^\mathrm T \mathcal B q + q ^\mathrm T \mathcal C_t ,
\end{align}
where ${\mathcal B  = \mathcal B ^\mathrm T}$ is time-independent.
The vector components ${q_\xi }$ are operators verifying the \ac{CCR}
\begin{align} \label{m_21}
    & \comm{  q_\xi}{ q_\eta} = - i \hbar  J _{\xi \eta} ,\quad \xi,\eta \in \{1,\dots, 2 N\}, 
\end{align}
where  ${ J = \begin{bmatrix} 0 & I \\ - I & 0 \end{bmatrix} }$ is the standard symplectic matrix in ${ \mathbb R^{2 N\times 2 N} }$, ${I}$ the identity in ${ \mathbb R^{N\times N} }$ and ${J^\mathrm T = J ^{-1} = -J }$.

The dynamics generated by the Hamiltonian in \cref{m_184} can be described in terms of the integrals of motion (which do not assume hermiticity of the Hamiltonian)~\cite{dodonov2003theory}
\begin{align} \label{m_23}
    & q (t) = U_t q U_t ^{-1} =  \Lambda_t q + \Delta_t ,
\end{align}
where 
\begin{align} \label{m_24}
    & \Lambda _t =  \exp {J \mathcal B t } , \quad \Lambda_0 = E := \begin{bmatrix} I & 0 \\ 0  & I \end{bmatrix} \\
    \label{m_106}
    & \dot \Delta_t  = \Lambda_t J \mathcal C_t, \quad \Delta_0 = 0 := \begin{bmatrix} 0_n & 0_n \end{bmatrix}^\mathrm T ,
\end{align}
and the matrix $\Lambda_t$ is symplectic for every time $t$, with ${ \Lambda_t ^{-1} = J \Lambda_t ^\mathrm T  J ^\mathrm T }$~\cite{gadella1995moyal}.
In the Heisenberg picture, the canonical operator reads~\cite{akhundova1982wigner,dodonov2003theory}
\begin{align} \label{m_160}
    & q_H (t) := U_t ^{-1}\, q\,  U_t = \Lambda^{-1} _t  ( q - \Delta _t) ,
\end{align}
whose expectation value is given by
\begin{align} \label{m_122}
    & \expval{  q} _t =\Lambda ^{-1}_t ( \expval{   q}_0 - \Delta _t) .
\end{align}
In the phase-space representation of quantum mechanics, \cref{m_23,m_24,m_106} fully specify the state of the system at any time~\cite{gadella1995moyal}.
There, the density matrix $\rho_t$ is represented by a distribution ${W_t (z)}$, known as the Wigner function, where the vector ${z\in \mathbb R^{2N}}$ lists the momenta and coordinates.
For an arbitrary initial state ${ W_0 ( z) }$, the unitary evolution resulting from a quadratic Hamiltonian (see \cref{m_184}) is such that~\cite{dodonov2003theory}
\begin{align} \label{m_27}
    &  W ( z,\, t) =  W_0 ( \Lambda _t  z + \Delta _t) .
\end{align}
In the phase space, the evolution governed by the \ac{QPM} Hamiltonian coincides with that of its classical counterpart (obtained by taking ${\hbar \to 0}$ in the \ac{CCR}), a property characteristic of quadratic Hamiltonians~\cite{gadella1995moyal}. 
As a result, all quantum information is encoded in the system's initial state. 
Notably, the key distinction between \ac{CPM} and \ac{QPM} lies in the latter's ability, at least in principle, to be initialized in a nonclassical state, such as a Schrödinger cat state~\cite{dodonov2003theory}.
However, this does not complicate the evolution, as the resulting dynamics remain independent of the initial state, as demonstrated by \cref{m_27}.

In addition, the class of Gaussian Wigner functions remains closed under evolutions generated by quadratic Hamiltonians.
Specifically, the Wigner function is given by
\begin{align} \label{m_26}
    & W ( z,t) = \abs {\det {2 \pi M_t }}^{-1/2}  \nonumber\\
    & \exp{ -\frac{1}{2} ( z - \expval{ q}_t )^\mathrm T M_t^{-1} ( z -\expval{ q}_t ) } ,
\end{align}
where  ${M_t := \Lambda_t ^{-1} M_0 (\Lambda_t ^{-1} )^\mathrm T}$.
For Gaussian states, the following identity can be proven~\cite{quintela2024quantum,de2017wigner}
\begin{align} \label{m_80}
    & \expval {\delta q (\delta q)^\mathrm T} _t  = M_t + \frac{i\hbar}{2} J^\mathrm T ,
\end{align}
with ${\delta q := q -\expval q_t}$.

Before proceeding with further calculations, it is important to note that the evolution of the operator ${q}$ under the symplectic matrix ${\Lambda_t = \exp {J \mathcal B t }}$ does not provide any new information beyond what was already known.
This can be demonstrated by examining the roots of the characteristic polynomial ${p_{J\mathcal B}(\lambda) = \det{J \mathcal B -\lambda E}=0}$, where  ${ J \mathcal B = \begin{bmatrix} 0 & A^{-1} \mathcal K \\ -A & 0 \end{bmatrix}}$.
These contain the relevant spectral information of ${\Lambda_t}$.
Using the formula for the determinant of a block matrix and assuming ${\lambda \neq 0}$, we find that ${p_{J\mathcal B}(\lambda) = \det{ \lambda + i \sqrt \mathcal K }\det{ \lambda - i \sqrt \mathcal K }}$, leading to ${\lambda \in Sp\{\pm i \sqrt \mathcal K \} \cup \{0\}}$ or, in a more compact form:
\begin{align} \label{m_352}
    &  Sp\{ J \mathcal B\} = \pm Sp \big \{ i \sqrt \mathcal K \big\},
\end{align}
where the zero can be omitted based on \cref{m_324}.
For a detailed derivation, refer to the Supplemental Material~\footnote{See Supplemental Material (Section III A) at \url{http://} for the spectral analysis of ${J \mathcal B}$.}.

\subsection{Self-Consistent Electric Field} \label{m_260}

We now apply the above theoretical results to the calculation of relevant physical quantities, starting with the electrical field.
When the \ac{QPM} is excited by an external electric field, either generated by a \ac{QM} system or an external source, the medium polarization sources are polarized and an electric field is re-emitted. 
This emitted field can, in turn, interact with the medium itself, generating a self-consistent interaction mechanism.
Depending on the structure of the medium, the emitted field can be enhanced or suppressed in specific regions, which is key to various nanoplasmonic applications~\cite{10.1021/acs.jpcc.5b09294,C3CS60364D,1530-6984,ZHOU20191,10.1021/acs.accounts.9b00308}. 

To determine the electric field emitted by the \ac{QPM}, we resort to the semiclassical approximation, where the quantum matter evolution and the classical Maxwell equation for the propagating field, assuming nonmagnetic interactions, are related through~\cite{mukamel1995principles}
\begin{align} \label{m_129}
    & \nabla \times \nabla \times \vec{ \mathcal E}(r,t ) + \frac{1}{c^2} \pdv{^2}{t^2} \vec{\mathcal E}(r,t ) = -\frac{4 \pi}{c^2} \pdv{^2}{t^2} \vec P(r,t ) \\
    \label{m_137}
    & \vec P(r,t ) = \tr{\vec V(r) \rho (t)} \\
    \label{m_144}
    & \dv{}{t}\rho(t) = - \frac{i}{\hbar} \comm{H(t)}{\rho (t)} \\
    \label{m_136}
    & H(t) = \frac{1}{2}q^\mathrm{T}\mathcal{B}q + H_\mathrm{int}(t).
\end{align}
Here, ${\vec P (r,t)}$ is the polarization, given as the expectation value of the dipole operator ${\vec V(r)}$, which, in components, is expressed as:
\begin{align} \label{m_196}
    & V_j (r) := \sum_{\alpha=1}^n R_{j \alpha} u _{\alpha} G (r;\, R_\alpha, \Sigma_\alpha).
\end{align}
This formula is obtained from the dipole approximation for the polarization~\cite{mukamel1995principles}, with an additional replacement of delta-like densities by the Gaussian functions ${G (r;\, R_\alpha, \Sigma_\alpha)}$. 
We have also identified the polarization sources ${u_\alpha (t)}$ with electric charges, and the generalized coordinate vector with the corresponding spatial coordinates.

The interaction of the charges ${u_\alpha (t)}$ with an external field is given by~\cite{mukamel1995principles}
\begin{align} \label{m_91}
    & H_1 (t) = - \sum_{j=1}^3 \int _{\mathbb R^3} \dd r\,  \mathcal E_j (r,\, t)  V_j (r) .
\end{align}
The Hamiltonian ${H_1(t)}$ determines, in turn, an interaction term ${H_2(t)}$ with the new variables ${v_\alpha (t)}$, resulting from the interaction term ${x^\mathrm T A^{-1} F(t)}$ and \cref{m_203}. 
More explicitly, we have the following relations 
\begin{align} \label{m_261}
    & F(t) = A \begin{bmatrix} h(t) \\ g(t) \end{bmatrix} \Leftrightarrow 
        \begin{bmatrix} f(t) \\ \dot f(t) - 2 \Gamma f(t) \end{bmatrix} 
            = \begin{bmatrix} A_1 & A_2 \\ A_2^\mathrm T & A_3 \end{bmatrix} \begin{bmatrix} h(t) \\ g(t) \end{bmatrix} \\
    \label{m_118}
    & h_\nu (t) := - \sum_{j=1}^3 R_{j \nu } \int_{\mathbb R ^3 } \dd r\, \mathcal E_j  (r, t) G (r; R_\nu, \Sigma_\nu ) ,
\end{align}
where ${A = \begin{bmatrix} A_1 & A_2 \\ A_2^\mathrm T & A_3 \end{bmatrix}}$, ${A_1 = A_1^\mathrm T}$, ${A_3 = A_3^\mathrm T}$ are given by \cref{m_181}, and ${ g (t) }$ is the unknown to be determined from the classical field ${ h (t)}$ and \cref{m_261} through the vector-form differential equation
\begin{align} \label{m_204}
    & A_2 \dot g(t) - (A_3 + 2 \Gamma A_2) g(t) = (A_2^\mathrm T + 2\Gamma A_1) h(t) - A_1 \dot h(t).
\end{align}
For a step-wise derivation of this equation see the Supplemental Material~\footnote{ See Supplemental Material (Section VIII) at \url{http://} for the derivation of the differential equation for the external field interacting with the new variables ${v_\alpha (t)}$.}.
\Cref{m_204} is formally solved in the frequency domain by
\begin{align} \label{m_211}
    & \tilde g (\omega ) =\tilde{  \mathcal L} (\omega ) \tilde  h (\omega )\\
    \label{m_188}
    &\tilde { \mathcal L} (\omega ) := - \big[ A_3 + (i \omega + 2 \Gamma ) A_2 \big] ^{-1} \big[ A_2^\mathrm T +  (i \omega + 2 \Gamma ) A_1 \big] ,
\end{align}
where ${\tilde g (\omega),\, \tilde h (\omega)}$ are the \ac{FT} of ${g(t),\, h(t)}$, with 
\begin{align}
    \label{m_125}
    &\tilde h_\nu (\omega ) = -\frac{1}{(2\pi)^3} \sum_{l=1}^3 R_{l \nu } \int_{\mathbb R ^3 } \dd k\,\tilde G (- k; R_\nu, \Sigma_\nu ) \tilde{\mathcal E}_l  (k, \omega)\\
    \label{m_127}
    & \tilde G (k; R_\alpha, \Sigma_\alpha ) = \exp{ -i k^\mathrm T R_\alpha - \frac{1}{2} k^\mathrm T \Sigma_\alpha k } .
\end{align}
Back to the time-domain, results the global Hamiltonian 
\begin{align} \label{m_117}
    & H_\mathrm{int} (t) = H_1(t) + H_2(t) = \sum_{\nu  = 1 }^n x_\nu h_\nu (t)  + x_{n + \nu} g_\nu (t) \\
    & g(t) = \mathcal L(t) \circledast h(t),
\end{align}
where ${\mathcal L (t)}$ is the \ac{FT} of ${\tilde {\mathcal L} (\omega)}$ and the symbol ${\circledast}$ indicates the convolution between the matrix ${\mathcal L (t)}$ and the vector ${h(t)}$, which is formally derived in the Supplemental Material~\footnote{ See Supplemental Material (Section VIII A) at \url{http://} for the definition and algebraic properties of the symbol ${\circledast}$.}.

The electric field is determined using the Green's function solution of Maxwell's equations in ${(k,\, \omega)}$-space~\cite{mukamel1995principles}
\begin{align} \label{m_61}
    &  \tilde{ \mathcal E}_i (k,\, \omega ) =  \tilde{  \mathcal E}_{ext,i} (k,\, \omega ) + \sum_{j=1}^3 \mathcal G_{ij} (k,\, \omega )  \tilde{P}_j (k, \omega ) \\
    \label{m_62}
    & \mathcal G_{ij} (k,\, \omega ) = 4 \pi \frac{\delta_{ij} \omega^2 - c^2 k_i k_j }{ \omega^2 - c^2 |k|^2 },\, k \in \mathbb R^3 .
\end{align}
Substituting \cref{m_196} into \cref{m_137} and taking the \ac{FT} results the polarization 
\begin{align} 
    \label{m_205}
        & \tilde{ P_j} (k,\omega ) = \sum_{\alpha =1}^n R_{j \alpha } \, \tilde G (k; R_\alpha, \Sigma_\alpha ) \expval q _{N+\alpha} (\omega ) ,
\end{align}
where the expectation value is taken over an arbitrary initial state (not necessarily Gaussian), resulting in the following analytical form (see the Supplemental Material~\footnote{ See Supplemental Material (Section IX A) at \url{http://} for the calculation of the polarization in the $k, \omega$ space.}):
\begin{align} 
    \label{m_126}
    & \expval q (\omega)  = 2 \pi \delta (\omega E + i J \mathcal B) \expval q_0 - i ( \omega E + i J \mathcal B )^{-1} J \tilde{\mathcal C} (\omega) .
\end{align}
In the above expression, all quantum information is encoded in the system's initial state and enters through the expectation value ${\expval q_0}$, while the electric field appears in the quantity ${ \tilde{ \mathcal C} (\omega)}$, given by
\begin{align} \label{m_198} 
    & \tilde{\mathcal C} (\omega ) = \begin{bmatrix} 0 & 0 & \tilde h(\omega) &  \tilde g(\omega) \end{bmatrix}^\mathrm T.
\end{align}
The self-consistent relation for the electric field ${\tilde{\mathcal E} }$ arises from its presence on both sides of \cref{m_61}, as can be traced through \cref{m_205,m_126,m_198} and the functional dependences ${\tilde h[\tilde{\mathcal E}],\, \tilde g[ \tilde{\mathcal E}]}$. 
Assuming that ${\expval q_0=0}$, we obtain the following expression:
\begin{widetext}
\begin{subequations} \label{m_262}
\begin{align} \label{m_212}
    & \tilde{ \mathcal E}_i (k,\, \omega ) =  \tilde{  \mathcal E}_{ext,i} (k,\, \omega ) + \sum_{j,l=1}^3 \sum_{\alpha =1}^n\mathcal G_{ij} (k,\, \omega )  R_{j \alpha } \, \tilde G (k; R_\alpha, \Sigma_\alpha ) \int_{\mathbb R^3} \dd k'\, T_{\alpha l} (-k', \omega ) \tilde{\mathcal E}_l (k', \omega )  \\
    \label{m_213}
    & T_{\alpha l} (k, \omega ) := \frac{i}{(2\pi)^3} \sum_{\nu=1}^N \sum_{\beta=1}^n \big (\omega E + i J \mathcal B \big )^{-1} _{N+\alpha, \nu} \big ( \delta _{\nu \beta } + \theta (\nu-n-1) \mathcal L _{\nu -n, \beta }(\omega) \big) R_{l\beta} \tilde G ( k; R_\beta, \Sigma_\beta) ,
\end{align}
\end{subequations}
\end{widetext}
where ${\theta (x)}$ is the Heaviside function.

An iterative solution to \cref{m_262} can be found by considering an external field of the form
\begin{align} \label{m_187}
    & \tilde{\mathcal E} _{ext,i} (k,\omega) = (2\pi)^3 \sum_p \mathcal A _{p,i} (\omega ) \delta (k - k_p) ,
\end{align}
for which, to first order, results 
\begin{align} \label{m_214}
    & (2\pi)^{-3}\, \tilde{\mathcal E}_i (k,\omega) = \sum_p \Big ( \mathcal A _{p,i} (\omega ) \delta (k - k_p) \nonumber\\
    & + \sum_{j,l=1}^3 \sum_{\alpha =1}^n\mathcal G_{ij} (k,\, \omega )  R_{j \alpha } \, \tilde G (k; R_\alpha, \Sigma_\alpha ) T_{\alpha l} (-k_p, \omega ) \mathcal A _{p,l} (\omega ) \Big ) . 
\end{align}
In light of this formula, we can study the spatial enhancement of the electric field in terms of the spreading in
${k}$-space (yielding localization in ${r}$-space) as resulting from the superposition of the Gaussians ${\tilde G ( k; R_{\alpha/\beta}, \Sigma_{\alpha/\beta})}$, mediated by the resonances of the matrix ${(\omega E + i J \mathcal B )^{-1}}$.

\section{Open quantum dynamics in the Quantum Polarizable Medium} \label{m_274}

We now aim to describe the open quantum dynamics of a \ac{QM} system embedded in an environment of dissipative quantum oscillators, here represented by the \ac{QPM}.
    We start by defining the global (system + environment) Hamiltonian. Since, in the Hamiltonian formulation of the media, we have introduced a set of auxiliary variables ${v_\alpha}$, we must account for the system's interaction with these.
Unlike what we did in \cref{m_273} for classical fields, we are now dealing with quantum operators, and there should not be a unique way of defining the interaction Hamiltonian.  
The approach we will follow here is to specify a quantum interaction that is consistent, in mean value, with the results derived for the classical fields.

Once the full Hamiltonian is specified, we will follow the microscopic derivation of the quantum master equation. 
A main drawback, compared to the standard approach, is that the environment Hamiltonian is non-Hermitian.
As a result, the correlation functions will lack the symmetries present in the Hermitian case, which arise from identifying the evolved operator ${q_\alpha (t)}$ with its Hermitian conjugate, as is done in standard derivations.

\subsection{The Interaction Hamiltonian} \label{m_249}

The total Hamiltonian can be written as 
\begin{align} \label{m_246}
    & H =  H_S + H_0 + H_\text{int}
\end{align}
where ${H_S}$ is the Hamiltonian of the \ac{QM} system, ${H_0 := \frac{1}{2}q^\mathrm{T}\mathcal{B}q}$ represents the isolated \ac{QPM} with ${\mathcal B}$ as defined in \cref{m_182}, and ${H_\text{int}}$ is the interaction Hamiltonian.
Following Ref.~\cite{guido2020open}, the interaction term is proposed of the form ${\sum_{\alpha=1}^n A_\alpha u_\alpha}$, where ${A_\alpha}$ is the electrostatic potential  generated by the \ac{QM} system, which interacts with the charges ${u_\alpha}$ at sites ${R_\alpha}$.
Similarly to the discussed for the classical field ${h(t)}$ (see \cref{m_260}), the potential ${A_\alpha}$ induces an interaction term with the auxiliary variables ${v_\alpha}$, resulting in a Hamiltonian contribution ${B_\alpha v_\alpha}$.
The total interaction term is thus given by
\begin{align} \label{m_266}
    & H_\text{int} = \sum_{\alpha=1}^n A_\alpha u_\alpha + B_\alpha v_\alpha, 
\end{align}
where ${B_\alpha \equiv B_\alpha [A]}$ is a function of all operators ${\{A_\alpha\}_{\alpha=1,\dots,\,n} }$, which represents the unknown to be determined.

To determine ${B_\alpha}$, we first express ${A_\alpha}$ as a time-dependent quantity in the interaction picture:
\begin{align} \label{m_267}
    & A_\alpha (t) = e^{ i H_S t/\hbar} A_\alpha e^{-i H_S t/\hbar} = \sum_\omega e^{-i \omega t/\hbar}  A_\alpha (\omega),
\end{align}
where 
\begin{align} \label{m_346}
    & A_\alpha = \sum_\omega A_\alpha  (\omega),\quad A_\alpha  (\omega) := \sum_{\varepsilon' - \varepsilon =\omega}\Pi_\varepsilon A_\alpha \Pi_{\varepsilon'},
\end{align}
and $\Pi_\varepsilon$ is the projector onto the eigenbasis of the \ac{QM} Hamiltonian.
In deriving \cref{m_267} we are using the identity ${e^{i/\hbar H_\text{S} t}  A_\alpha (\omega) e^{-i/\hbar H_\text{S} t} = e^{-i/\hbar \omega t}  A_\alpha (\omega) }$.
Next, we proceed to an important step, where the results derived for the classical field in \cref{m_260} are applied:
identifying the classical field ${h_\alpha (t)}$ with the mean value of the operator ${A_\alpha}$ over the evolved \ac{QM} state ${e^{-i H_S t/\hbar} \rho_S e^{i H_S t/\hbar}}$, namely 
\begin{align} \label{m_363}
    & h_\alpha (t) = \tr{ A_\alpha \rho_S (t)} = \expval{ A_\alpha (t)} ,
\end{align}
follows that ${ \expval{ B_\nu (t)} = g_\nu (t) }$, as these terms correspond to the interaction with ${u_\alpha}$ and ${v_\nu}$, respectively.
Then, in the frequency domain
\begin{align} \label{m_364}
    & \expval{B(\omega)} = \tilde{\mathcal L} (\omega) \expval{A(\omega)},
\end{align}
where ${ \tilde{\mathcal L}(\omega)}$ is given by \cref{m_188}. 
As the trace is a linear operation and ${\rho_S}$ is arbitrary, this equality implies 
\begin{align} \label{m_264}
    &  B_\alpha (\omega) = \sum_\nu \tilde{\mathcal L}_{\alpha \nu} (\omega) A_\nu (\omega).
\end{align}
By summing over all frequencies, we obtain the operator
\begin{align} \label{m_265}
    & B_\alpha = \sum_{\omega,\, \nu} \tilde{\mathcal L}_{\alpha \nu} (\omega) A_\nu (\omega),
\end{align}
which, together with ${A_\alpha}$, completely specifies the interaction Hamiltonian ${H_\text{int}}$ in \cref{m_266}.

\subsection{Master Equation for the \ac{QM} System} \label{m_243}

The Markovian dynamics of the \ac{QM} system embedded in a \ac{QPM} is governed by the Master equation, which in the interaction picture reads~\cite{breuer2002theory}
\begin{align} \label{m_46}
    & \dv{}{t}\rho_S (t) = -\frac{i}{\hbar} \comm{H_{LS}}{\rho_S (t)} + \mathcal D (\rho_S (t)) .
\end{align}
Here, ${\rho_S (t)}$ is the \ac{QM} density matrix, ${H_{LS}}$ is the Lamb shift, and ${\mathcal D(.)}$ is the dissipator.
The dissipator is a superoperator that accounts for the decoherence in the system ${\rho_S}$ resulting from the environment---i.e., the interaction with the \ac{QPM}---and it is given by
{\small
\begin{align} \label{m_42}
    & \mathcal D (\rho_S) = \nonumber\\
    & \hbar^{-2}\sum_{\omega} \sum_{\alpha, \beta =1}^{N} \gamma_{\alpha\beta } (\omega) \Big(  O_\beta (\omega) \rho_S O^\dagger _\alpha (\omega)  -\frac{1}{2} \acomm{O_\alpha ^\dagger (\omega) O_\beta (\omega) }{ \rho_S} \Big) ,
\end{align}
}
where the sum runs over ${N=2n}$ variables of the \ac{QPM}, namely ${u_\alpha}$ and ${v_\alpha}$.
The system operators ${O_\alpha (\omega)}$, obtained from the interaction Hamiltonian ${H_\text{int}}$ in \cref{m_266}, are given by: 
\begin{align} \label{m_347}
    & O_{\alpha }(\omega) :=\begin{cases} A_\alpha(\omega) & \alpha \leq n \\ B_{\alpha -n } (\omega) & n < \alpha \leq 2n \end{cases},
\end{align}
where ${B_\alpha (\omega)}$ is given in \cref{m_264}.
We remind that, while ${A_\alpha}$ accounts for the interaction of the system with the variables ${u_\alpha}$, ${B_\alpha}$ describes the interaction with auxiliary variables ${v_\alpha}$.

The functions ${\gamma_{\alpha \beta}(\omega)}$ appearing in the dissipator are referred to as \textit{correlation functions} and are defined in terms of the operators ${x \coloneqq \begin{bmatrix} u & v \end{bmatrix}^\mathrm T}$ through the equations:
\begin{align} \label{m_43}
    & \gamma_{\alpha \beta }(\omega) := \Xi_{\alpha \beta } (\omega) + \Xi^*_{\beta \alpha } (\omega) \\
    \label{m_361}
    & \Xi_{\alpha \beta } (\omega) := \int_{ \mathbb R \geq 0} \dd t\, \expval{ x_\alpha (t) x_\beta (0) } \exp (i t \omega ) ,
\end{align}
where
\begin{align} \label{m_161}
    & x_\alpha (t) = e^{i H_0 t /\hbar } x _\alpha e^{-i H_0 t /\hbar }, \quad 1\leq \alpha \leq 2n. 
\end{align}
In the microscopic derivation of the master equation (see Ref.~\cite{breuer2002theory}), the function ${\Xi_{\alpha \beta } (\omega)}$ is defined with ${x^\dagger_\alpha(t)}$ in place of ${x_\alpha (t)}$, which, together with the hermiticity of ${x_\alpha (t)}$, allows the correlation function to be written as the integral
    \begin{align} \label{m_365}
    & \gamma_{\alpha \beta }(\omega)  = \int_{ \mathbb R } \dd s\, \expval{ x_\alpha^\dagger (s) x_\beta (0) } \exp (i s \omega) .
    \end{align}
    This is not be possible here, as ${x_\alpha (t) \neq x_\alpha^\dagger (t)}$ due to the non-Hermiticity of the \ac{QPM} Hamiltonian.

The Lamb shift ${H_{LS}}$ consists of a Hermitian operator containing the environment-induced corrections to the system energies, and satisfies the identity ${\comm{H_S}{H_{LS}} = 0}$~\cite{breuer2002theory}.
It is given by the expression:
\begin{align} \label{m_359}
    & H_{LS} = \hbar^{-1} \sum_\omega \sum_{\alpha, \beta=1}^{2n} S_{\alpha \beta } (\omega) O_\alpha ^\dagger (\omega) O_\beta (\omega) ,
\end{align}
where
\begin{align} \label{m_360}
    & S_{\alpha \beta} (\omega) := \frac{1}{2i} \big( \Xi_{\alpha \beta } (\omega) -\Xi^*_{\beta \alpha } (\omega) \big) .
\end{align}
We now aim computing the matrix elements ${ \Xi_{\alpha \beta} (\omega) }$, from which it can be calculated the correlation functions ${\gamma_{\alpha \beta} (\omega) }$ and the coefficients ${S_{\alpha \beta}(\omega)}$, via \cref{m_43,m_360}.
We assume that the \ac{QPM} is initialized in a Gaussian state ${\rho_G}$ with first moments ${\tr{q_\alpha (t) \rho_G} = 0}$ and under no external field, such that ${\Delta_t = 0}$ (see \cref{m_106}).
From \cref{m_161,m_160,m_43,m_80}, we have 
\begin{align} \label{m_141}
    & \Xi (t)\coloneqq \expval{ q_H (t) q^\mathrm T }_0 \nonumber\\
    & = \exp{- J \mathcal B t} \Big (M_0 + \frac{i \hbar}{2} J^\mathrm T + \expval q_0 \expval q_0^\mathrm T \Big ),
\end{align}
where this compact matrix notation allows the indices ${\alpha,\, \beta}$ of ${\Xi_{\alpha \beta }(t)}$ to run over the entire interval ${\alpha,\,\beta =1,\dots, 2N}$, which is permitted for notational convenience, even though we are only concerned with ${\alpha,\, \beta = N+1,\dots 2N}$, since ${q \coloneqq \begin{bmatrix} \pi_u & \pi_v & u & v \end{bmatrix}^\mathrm T}$.
        Note that the function ${\Xi (t)}$ explicitly depends on the eigenvalues ${ Sp\{J \mathcal B\}= Sp\{\pm \sqrt \mathcal K \} }$, to which the matrices ${K}$ and ${\Gamma}$ contribute via \cref{m_215}.

To compute the half-domain \ac{FT} appearing in \cref{m_361}, we use the identity
\begin{align} \label{m_362}
    &  \lim_{\eta \to 0^+ } \int _0 ^\infty \dd t\, e^{ i ( \Omega + i \eta ) t} = i \lim_{\eta \to 0^+ } (\Omega + i \eta )^{-1} 
\end{align}
which, upon substituting ${\Omega = \omega E + i J \mathcal B}$, gives:
\begin{align} \label{m_285}
    & \Xi (\omega) = \nonumber\\
    & i \lim_{\eta \to 0^+} \Big( (\omega + i \eta) E + i J \mathcal B  \Big)^{-1} \Big (M_0 + \frac{i \hbar}{2} J^\mathrm T + \expval q_0 \expval q_0^\mathrm T \Big ).
\end{align}
For an initial thermal state in absence of an external field, the covariance matrix and mean vector are given by~\cite{akhundova1982wigner}
\begin{align} \label{m_77}
    & M_0 = -\frac{\hbar}{2} \cot \Big (  \frac{\hbar \beta J \mathcal B }{2} \Big) J ^\mathrm T,\quad \expval q _0 = 0.
\end{align}
Using the identities ${\cot x = i \coth (i x )}$ and ${J^\mathrm T = - J}$, we find
{\small
\begin{align} \label{m_216}
    & \Xi (\omega) = -\frac{\hbar}{2} \lim_{\eta \to 0^+} \Big( (\omega + i \eta) E + i J \mathcal B  \Big)^{-1} \Big( \coth ( \frac{\hbar \beta i J \mathcal B}{2}) -1 \Big) J .
\end{align}
}
Expressing the hyperbolic cotangent function in terms of the Bose-Einstein distribution ${n_{BE} (x) \coloneqq (e^x -1)^{-1}}$, namely ${\coth \big( \frac{x}{2} \big)  = 2 n_{BE} (x) +1}$, results
\begin{align} \label{m_366}
    & \Xi (\omega) = -\hbar \lim_{\eta \to 0^+} \Big( (\omega + i \eta) E + i J \mathcal B  \Big)^{-1}  n_{BE} (\hbar \beta i J \mathcal B) J .
\end{align}
Assuming that ${\mathcal B}$ is independent of $\hbar$, meaning that the matrices $K$ and $\Gamma$ in \cref{m_185} do not contain scaled quantum corrections, the classical limit of the function ${\Xi(\omega)}$ can be evaluated from the identity ${\lim_{x\to 0} \frac{x}{e^{a x}-1}= \frac{1}{a}}$, resulting in
\begin{align} \label{m_82}
    & \lim_{\hbar \to 0} \Xi (\omega) = - (\beta i J \mathcal B)^{-1} \lim_{\eta \to 0^+} \Big( (\omega + i \eta) E + i J \mathcal B  \Big)^{-1} J.
\end{align}
Quantum corrections to this function can be accounted for perturbatively by expanding the Bose-Einstein distribution in \cref{m_366} into a power series, thus yielding a generalization of the results presented in Ref.~\cite{guido2020open}.

\section{Conclusions} \label{m_275}

In conclusion, we have presented a comprehensive theoretical framework for understanding the classical and quantum dynamics of polarizable media and their interaction with quantum systems. 
Our approach provides a unified perspective by incorporating key concepts from pseudo-boson theory, phase-space formulation, and open quantum systems, with potential applications in condensed matter physics and nanoplasmonic materials.

We began by presenting the theoretical foundations, demonstrating how the \ac{CPM} defining equation can be integrated in an expanded space, where the spectral properties of the dynamics are encapsulated by a time-independent matrix, ${\sqrt \mathcal K}$.
The formulation then introduces a quadratic Hamiltonian, quantized under the condition that its mean-field dynamics coincide with those of the \ac{CPM}, thereby defining the \ac{QPM}.
This Hamiltonian was shown to yield a conserved quantity for the isolated system (${f(t)=0}$) and was subsequently diagonalized in terms of pseudo-boson operators. 
The diagonalization enabled us to express the dynamics over Bi-Coherent states, providing a deeper understanding of the system's behavior.

Building on these foundations, we applied our theoretical results to the analysis of polarization signal measurements. 
By deriving explicit formulas for decomposing the polarization in terms of the eigenvalues and eigenvectors of ${\sqrt \mathcal K}$, we established a method to organize and quantify their contributions to the spectral properties of the polarizable medium. 
This analysis provides valuable insights into the spectral structure of these systems, as well as the role of individual eigenvalues in defining their response to external fields.

We then applied our formalism to derive the (semi-classical) electric field emitted by the polarizable medium following an electric excitation. 
The \ac{QPM} dynamics was formulated in phase space, offering a classical analogy while enabling the analytical calculation of expectation values over arbitrary \ac{GS}, including thermal states. 
Assuming interaction with a classical field, we derived a self-consistent equation for the emitted electric field in frequency-momentum space. 
This formulation led to a perturbative solution, whose analytical properties can be exploited to understand electric field enhancements in nanoplasmonic materials.

Finally, we formulated the open quantum dynamics of a \ac{QM} system interacting with the \ac{QPM}. 
Using the theoretical results obtained for classical electric fields, we derived the interaction Hamiltonian for the original and auxiliary variables, everything of which contribute to the quantum master equation. 
We provided analytical expressions for the \ac{QPM} correlation functions, which are valid for arbitrary initial Gaussian states.

The combination of classical polarizable approaches with the quantum mechanical formulation highlights the versatility of this framework. 
These results not only generalize existing approaches but also open the door to novel applications across materials science, nanoplasmonics, condensed matter, and quantum technologies. 
For instance, the explicit solutions for electric fields could pave the way for advances in light-matter interaction, while the quantum master equation introduces new avenues for modeling dissipative quantum systems embedded in \ac{QPM}. 
The correlation functions derived for \ac{GS} offer analytical tools for exploring thermal effects, time evolution, and other phenomena relevant to quantum systems embedded in complex polarizable environments. 
Future research could also extend this framework to explore strong coupling regimes between a \ac{QM} system and the \ac{QPM}.

\section*{Acknowledgments}
The authors acknowledge funding from MUR-FARE Ricerca in Italia: Framework per l'attrazione ed il rafforzamento delle eccellenze per la Ricerca in Italia - III edizione. Prot. R20YTA2BKZ.
The authors also thank Vasco Cavina for informative and insightful discussions. 
The authors declare no conflicts of interest to disclose.

\section*{Acronyms}
\begin{acronym}
    \acro{CCR}{Canonical Commutation Relation}
    \acro{CPM}{Classical Polarizable Medium}  
    \acro{EF}{Eigenvalue Filter}
    \acro{FT}{Fourier Transform}
    \acro{GS}{Gaussian States}
    \acro{IF}{Intercept Filter}
    \acro{QM}{Quantum Mechanics}
    \acro{QPM}{Quantum Polarizable Medium}
\end{acronym}

\bibliography{bib}

\appendix
\onecolumngrid

\tableofcontents
\makeatletter
\let\toc@pre\relax
\let\toc@post\relax
\makeatother

\section{Equations of Motion for the Classical Polarizable Medium (CPM)} \label{a_149}

\subsection{Non-dissipative CPM}

\subsubsection{Fluctuating Charges/Fluctuating Dipoles force fields} \label{a_202}

Let us consider a set of $n$ atoms, located at (fixed) positions $R_i$ with $i=1,\dots,n$. To simulate the polarization of this system to the electromagnetic field, the Fluctuating Charges (\fq) force field represents each atom as a charge $q_i(t)$ that is allowed to vary as a response to the external field. Furthermore, we want to divide the $n$ atoms in $m$ fragments and impose that the total charge on each fragment is constrained to a certain value $c_\alpha$, with $\alpha=1,\dots,m$. 
Under these assumptions, the Lagrangian functional of the system can be written as~\cite{rick1994dynamical,rick1995fluctuating}:
\begin{align}\label{eq:lagrangian-fq}
    & \mathcal{L}(q,\dot{q},\lambda) = \frac{1}{2} \dot q^\mathrm T M_q \dot{q} - \frac{1}{2} q^\mathrm T \tqq q - q^\mathrm{T}(\chi+v(t)) - \lambda^\mathrm T (c - 1_\lambda ^\mathrm T q), 
\end{align}
where $M_q$ is the mass matrix, $\tqq$ is the charge-charge interaction kernel, $\chi_i$ is the electronegativity of the atom $i$, and $v_i(t)$ is the external electric potential acting on the same site. $1_\lambda$ is a ${n\times m}$ rectangular matrix defined as
\begin{align}
    & (1_\lambda)_{i\alpha} = 
    \begin{cases}
    1 & \mbox{if atom } i \in \mbox{ fragment }\alpha \\
    0 & \mbox{otherwise} .
    \end{cases}
\end{align}
$\tqq$ is a function of the interatomic distance ${R_{ij} = \absu{R_i-R_j}}$ and it describes the electrostatic interaction between the charges.
Due to the divergence of the standard Coulomb interaction for ${R_{ij}\rightarrow 0}$, charge-charge interactions may become too strong and lead to the so-called ``polarization catastrophe''~\cite{thole1981molecular} in which the charges are extremely large, due to the classical description of purely quantum quantities. 
For this reason, alternative \textit{damped} formulations of $\tqq$ have been proposed, characterized by the convergence to a finite limit with ${R_{ij}\rightarrow 0}$~\cite{ohno1964some,mayer2007formulation}.
This is usually realised by assuming that the charges $q_i$ are associated with a spherical distribution ${\rho(R)}$ centered around the position $R_i$.
From the formal point of view, $\tqq$ is a symmetric, positive definite matrix.
The dynamical equation for the \fq model reads
\begin{align}\label{eq:eom-fq}
    & M_q\ddot{q} = \tqq q + \chi + v(t) .
\end{align}
The \fq force field has been recently extended to simulate a more refined polarization employing electric dipoles. In the resulting Fluctuating Charges and Fluctuating Dipoles (\fqfmu) force field, each atom is endowed with a charge $q_i$ and an electric dipole $\mu_i$ describing the polarization of the system as induced by the external electromagnetic field~\cite{giovannini2019polarizable}.
The dynamical equation for the \fqfmu model reads
\begin{align} \label{eq:eom-fqfmu}
    & M_q\ddot{q}      = \tqq q + \tqmu \mu + \chi + v(t) \\
    & M_\mu\ddot{\mu}  = \tmuq q + \tmumu \mu + e(t),
\end{align}
where $\tqmu,\tmumu$ are the (damped) charge-dipole and dipole-dipole interaction kernels~\cite{mayer2007formulation}, while $e_i(t)$ is the external electric field acting on the atom $i$.

\subsubsection{Induced dipole force fields}

Another important family of classical force fields is represented by the induced dipole models. Here, the system is represented as the sum of a permanent electrostatic distribution (up to the octupole $\Xi$ depending on the model) and a time-dependent induced dipole $\mu$, both located on each atom of the molecular system. Several models belong to this family: Induced Point Dipole (IPD or MMPol)~\cite{thole1981molecular,thompson1996qm}, Polarizable Embedding (PE)~\cite{olsen2010excited}, Discrete Reaction Field (DRF)~\cite{jensen2003discrete}, Atomic Multipole Optimized Energetics for Biomolecular Applications (AMOEBA)~\cite{ponder2010current}. Similarly to the \fq and \fqfmu force fields, the equation of motion of the induced dipole moments read:
\begin{align} \label{eq:eom-fqfmu}
    & M_\mu\ddot{\mu}  = \tmumu \mu + e_\mathrm{static} + e(t),
\end{align}
where $e_\mathrm{static}$ is the electric field generated by the static multipolar distribution. 

\subsubsection{Continuum models}

Continuum embedding force fields are the most adopted methods in the context of multiscale simulations. Here, the system is represented as a continuum infinite dielectric solely described by its dielectric permittivity $\varepsilon$ and by the shape of the cavity in which usually a target molecule can be located. The response of the system to the external electromagnetic field is represented by a surface charge density $\sigma(r)$. The surface of the cavity is discretized in \textit{tesserae} and a point charge $q$ is located on the center of each \textit{tessera} mimicking the surface charge density. Under the assumption that the optical response of the system can be described in a time-independent fashion using the infinite-frequency permittivity $\varepsilon_\infty$, it is possible to derive an equation of motion for the charges~\cite{pipolo2014cavity}
\begin{equation}
M_q\ddot{q} = \tqq q + v(t)
\end{equation}
where the specific shape of $M_q$ and $\tqq$ depends on the formulation of the continuum method, such as the COnductor-like Screening MOdel (COSMO)~\cite{klamt1993cosmo} or the Integral Equation Formalism of the Polarizable Continuum Model (IEFPCM)~\cite{cances1997new,tomasi2005quantum}.

\subsection{Inclusion of decoherence}

\subsubsection{Atomistic CPM}

The FQ force field has been recently extended to describe the optical response of plasmonic substrates under the action of an external oscillating electric field. The resulting model is called the Frequency-Dependent Fluctuating Charges (\wfq) model, and it has been applied to the simulation of alkali metals nanostructures~\cite{giovannini2019classical} and graphene-based materials~\cite{giovannini2020graphene}.
In the frequency domain, each atom of the substrate is endowed with a frequency-dependent charge $q(\omega)$ and the optical response of the system is modeled by resorting to the Drude model of conduction, and by including a phenomenological description of quantum tunneling between the atomic sites. A real-time version of such model has been recently proposed, and the equation of motion reads~\cite{lafiosca2024real}:
\begin{equation}\label{eq:eom-wfq}
M_q\ddot{q} + M_q\Gamma_q\dot{q}= K^\mathrm{D-T}\tqq q + K^\mathrm{D-T}v(t), 
\end{equation}
where ${K^\mathrm{D-T}}$ is a matrix accounting for both Drude and tunneling mechanisms~\cite{lafiosca2021going}, while $M$ and $\Gamma$ are diagonal matrices whose functional form depends on the Drude parameters entering the model. 

Similar to the \fq model, the \fqfmu model can be generalized to define the \wfqfmu model, which is particularly suitable for the description of noble metal nanoparticles~\cite{giovannini2022we}.
In this model, the intraband contribution to the optical response is described by the frequency-dependent charges $q_i(t)$, while the frequency-dependent dipoles $\mu_i(t)$ take into account the effect of interband transitions. This is done by extracting the interband polarizability from the frequency-dependent permittivity of the material $\varepsilon(\omega)$, which is then expressed as the superposition of Drude-Lorentz oscillators (DLO)~\cite{lafiosca2024real}.
The dynamical evolution of charges and dipoles in the \wfqfmu model is given by~\cite{lafiosca2024real}:
\begin{align} \label{eq:eom-wfqfmu}
    & M_q\ddot{q} + M_q\Gamma_q\dot{q} = K^\mathrm{D-T}\tqq q + K^\mathrm{D-T}\tqmu \mu + K^\mathrm{D-T}v(t) \\ 
    & M_{\mu,p}\ddot{\mu}_p + M_{\mu,p}\Gamma_{\mu,p}\dot{\mu}_p = \tmuq q + \sum_q\left(\tmumu - \delta_{pq}N_{\mu,p}\right)\mu_q + e(t) ,
\end{align}
where $p$ is an index running over the DLO and $M_{\mu,p}, \Gamma_{\mu,p}$ and $N'_{\mu p}$ are defined in terms of the DLO parameters. A similar expression to \cref{eq:eom-wfqfmu} can in principle be obtained also for the Discrete Interaction Model (DIM), in which the optical response is described in terms of electric dipoles $\mu(\omega)$ ruled by the frequency-dependent atomic polarizability $\alpha(\omega)$~\cite{jensen2009atomistic}.

\subsubsection{Continuum CPM}

The Time-Domain Boundary Element Method (TD-BEM) is a continuum approach to the description of the time-dependent polarization of nanoparticles~\cite{pipolo2016real,dall2020real}.
In this method, the plasmonic substrate is modeled as a continuum medium whose optical response is ruled by $\varepsilon(\omega)$. In accordance with the non-dissipative methods such as COSMO and IEFPCM, the response to the electromagnetic field is modeled as a frequency-dependent surface charge density $\sigma(r,\omega)$, which undergoes a discretization process as point charges located on the center of each \textit{tessera} of the surface. Similar to the \wfqfmu model, $\varepsilon(\omega)$ is fitted as a superposition of DLO, and the equation of motion for the charges $q(t)$ becomes~\cite{dall2020real}:
\begin{align}\label{eq:eom-bem}
    & M'_{q,p}\ddot{q} + M'_{q,p}\Gamma'_{q,p}\dot{q}_p= \sum_q\left(D - \delta_{pq} N'_p \right)q_p + r_p(t) ,
\end{align}
where, in accordance with \cref{eq:eom-wfqfmu}, $M'_{q,p}, \Gamma'_{q,p}$ and $N'_{q p}$ are defined in terms of the DLO parameters. 

\section{ Hamiltonian in the expanding coordinates } \label{a_01}

Here we present a version of the expanding coordinates Hamiltonian. 
The Hamiltonian is constrained to be of the quadratic form 
\begin{align} \label{a_03}
    &  H = \frac{1}{2} \pi^\mathrm T A(t) \pi  + \frac{1}{2} u^\mathrm T C(t) u  + \pi^\mathrm T B(t) u + u^\mathrm T A(t) ^{-1} f(t) , 
\end{align}
with ${C(t) = C(t)^\mathrm T}$ and ${A(t) = A(t)^\mathrm T}$ invertible. 
In what follows, we will often omit the explicit time dependence or include the time as a subscript ${t}$ to ease the notation.

Applying the Hamilton equations~\cite{goldstein2002classical}
\begin{align} \label{a_04}
    & \dot u = \pdv{H}{\pi}, \quad  \dot \pi = - \pdv{H}{u} 
\end{align}
result in the differential equation 
\begin{align} \label{a_05}
    & \ddot u(t)  + 2 \Gamma \dot u(t) + K u(t) + f(t) = 0 , 
\end{align}
where 
\begin{align} \label{a_06}
    & K = \dot A_t A_t^{-1} B_t + A_t C_t - A_t B_t^\mathrm T A_t^{-1} B_t - \dot B_t \\
    \label{a_10}
    & 2 \Gamma =   A_t B_t^\mathrm T A_t^{-1} - B_t - \dot A_t A_t^{-1},
\end{align}
and ${K,\, \Gamma}$ will be assumed to be time-independent.
Following the \wfq formulation~\cite{giovannini2019classical}, we take ${\Gamma = \gamma I}$ and $K$ to be an arbitrary matrix,with no assumptions made about its symmetry or invertibility.
Choosing 
\begin{align} \label{a_07}
    & C_t = B_t^\mathrm T A_t ^{-1} B_t
\end{align}
we verify ${C_t = C_t^\mathrm T}$ and reduce \cref{a_06} to 
\begin{align} \label{a_08}
    & K = \dot A_t A_t^{-1} B_t - \dot B_t .
\end{align}
We can then propose the solutions
\begin{align} \label{a_09}
    & B = b_t K, \quad A = a_t S, 
\end{align}
where the time-dependence is captured by an scalar, and ${S = S^\mathrm T}$ is an invertible matrix. 
Substituting \cref{a_09} in \cref{a_08,a_10} results
\begin{align} \label{a_11}
    &  K = (\dot a a^{-1} b - \dot b) K \\
    \label{a_12}
    &  2 \gamma I = -  \dot a a^{-1} I + b ( S K ^\mathrm T S ^{-1} - K ) .
\end{align}
Now, for any $K$ there is an invertible ${\mathcal S= \mathcal S^\mathrm T}$ that achieves the similarity transformation~\cite{taussky1959similarity}
\begin{align} \label{a_13}
    & K = \mathcal S K^\mathrm T \mathcal S^{-1}.
\end{align}
Then, upon identifying ${S= \mathcal S }$ in \cref{a_12}, the second summand in \cref{a_12} vanishes, leaving us with the differential equations 
\begin{align} \label{a_14}
    & \dot a_t a_t^{-1} = - 2 \gamma , \quad \dot a_t a_t^{-1} b_t - \dot b_t  =1, 
\end{align}
which are solved by 
\begin{align} \label{a_15}
    & a_t = e^{- 2 \gamma t}, \quad b_t = e^{- 2 \gamma t} - \frac{1}{2 \gamma} . 
\end{align}
Collecting \cref{a_07,a_13,a_15}, we can write the Hamiltonian in \cref{a_03} as 
\begin{align} \label{a_153}
    &  H = \frac{1}{2} \pi^\mathrm T e^{- 2 \gamma t} S \pi  + \frac{1}{2} u^\mathrm T \Big ( e^{- 2 \gamma t} - \frac{1}{2 \gamma} \Big )^2 e^{ 2\gamma t} K^\mathrm T S^{-1} K  u  + \pi^\mathrm T K \Big ( e^{-2 \gamma t} - \frac{1}{ 2\gamma} \Big ) u + u ^\mathrm T e^{ 2 \gamma t} S^{-1} f_t .
\end{align}
Defining the expanding coordinates as 
\begin{align} \label{a_151}
    & x := \Big ( e^{- 2\gamma t} - \frac{1}{2 \gamma} \Big ) e^{ \gamma t} u \\
    & p := e^{- \gamma t} \pi ,
\end{align}
we can write the Hamiltonian from \cref{a_03} as 
\begin{align} \label{a_152}
    &  H = \frac{1}{2} p^\mathrm T S p  + \frac{1}{2} x^\mathrm T K^\mathrm T S^{-1} K x  + p^\mathrm T K x - x^\mathrm T e^{ \gamma t} \Big ( e^{ - 2 \gamma t} - \frac{1}{2 \gamma} \Big )^{-1} S^{-1} f_t , 
\end{align}
where the explicit time dependence is isolated in the term containing the field ${f_t}$.

\section{Classical Dynamics} \label{a_132}

We start with 
\begin{align} \label{a_133}
    & \ddot u + 2 \Gamma \dot u + K u + f_t = 0
\end{align}
which, under the variables change ${v = \dot u }$, gives
\begin{align} \label{a_154}
    & \ddot u + K u + 2 \Gamma v + f_t = 0.
\end{align}
To find the equation of motion for ${v}$, we take the derivatives
\begin{align} \label{a_137}
    & \dot v = \ddot u = - K u - 2 \Gamma v - f_t \\
    \label{a_136}
    & \ddot v = - K \dot u - 2 \Gamma \dot v - \dot f_t = -  K v  + 2 \Gamma K u + (2 \Gamma )^2 v +  (2 \Gamma f_t - \dot f_t ) .
\end{align}
Grouping the above results we get
\begin{align} \label{a_134}
    & \ddot u + K u + 2 \Gamma v + f_t = 0 \\
    & \ddot v + ( K - (2 \Gamma )^2 ) v - 2 \Gamma K u  + (\dot f_t - 2 \Gamma f_t ) = 0, \quad v(0)= \dot u(0),\, \dot v(0)= - K u(0) - 2 \Gamma \dot u(0), 
\end{align}
or in matrix notation 
{\small 
\begin{align} \label{a_138}
    & \dv[2]{t} \begin{bmatrix} u \\ v \end{bmatrix} 
        + \begin{bmatrix} K & 2\Gamma  \\ - 2 \Gamma K  &  K  - (2 \Gamma )^2 \end{bmatrix} \begin{bmatrix} u \\ v \end{bmatrix} 
        + \begin{bmatrix} f_t \\  \dot f_t - 2 \Gamma f_t \end{bmatrix} = \begin{bmatrix} 0 \\ 0 \end{bmatrix}.
\end{align}
}
Defining 
    \begin{align} \label{a_185}
        & x := \begin{bmatrix} u & v \end{bmatrix}^\mathrm T \\
        & \mathcal K := \begin{bmatrix} K & 2\Gamma  \\ - 2 \Gamma K  &  K  - (2 \Gamma )^2 \end{bmatrix} \begin{bmatrix} u \\ v \end{bmatrix} \\
            \label{a_173}
        & F(t) := \begin{bmatrix} f(t) &  \dot f(t) - 2 \Gamma f(t) \end{bmatrix}^\mathrm T,
    \end{align}
    we have  the velocity independent equation 
    \begin{align} \label{a_156}
        & \ddot x + \mathcal K x + F(t) = 0 , 
    \end{align}
    subject to the initial conditions 
    \begin{subequations} \label{a_237}
    \begin{align} 
        & x(0) = \begin{bmatrix} u(0) & \dot u(0) \end{bmatrix}^\mathrm T \\
        & \dot x(0) = \begin{bmatrix} \dot u(0) & - K u(0) - 2\Gamma \dot u (0) \end{bmatrix}^\mathrm T.
    \end{align}
    \end{subequations}
The homogeneous solution is constructed as a linear combination of the exponential functions  
\begin{align} \label{a_223}
    & \exp{\pm i \sqrt {\mathcal K}t},\quad \sqrt{\mathcal K} = i \begin{bmatrix} 0 & - I \\  K  &  2\Gamma \end{bmatrix} ,
\end{align}
where the coefficients are determined from the initial conditions \cref{a_237}.

\subsection{Spectral Analysis} \label{a_209}

Let ${Sp \{ M \} }$ denote the set of eigenvalues of the matrix ${M}$. 
We aim to show that
\begin{align} \label{a_217}
    & Sp \big \{ - \sqrt \mathcal K \big\} \subseteq \big\{ \omega : \det( \omega^2 + 2 i \omega \Gamma - K)=0 \big\} \cup \{0\},
\end{align}
which up to the constant solution (${\omega =0}$) correspond to those of \cref{a_133}.

\begin{proof}
This can be seen by noting that if ${A}$ is invertible, we have
\begin{align} \label{a_205}
    & \det \begin{bmatrix} A & B \\  C  &  D \end{bmatrix} = \det A \det (D - C A ^{-1} B).
\end{align}
The characteristic polynomial for ${\sqrt \mathcal K}$ reads
\begin{align} \label{a_206}
    & p_{\sqrt \mathcal K} (\lambda ) = \det{\sqrt{\mathcal K}-\lambda \text{Id}  } = \det{i \begin{bmatrix} 0 & - I \\  K  &  2\Gamma \end{bmatrix} - \begin{bmatrix} \lambda & 0 \\  0 &  \lambda \end{bmatrix}} = \det{ \begin{bmatrix} -\lambda & - i I \\  i K  &  2 i \Gamma -\lambda \end{bmatrix} }= 0 , 
\end{align}
where `Id' the identity matrix in ${ \mathbb R^{2n\times 2n}}$.
Assuming ${\lambda \neq 0}$ we can use the identity in \cref{a_205} to obtain
\begin{align} \label{a_207}
    & p_{\sqrt \mathcal K} (\lambda ) = \det{-\lambda I } \det{2 i \Gamma -\lambda  - i K ( -\lambda)^{-1} (- i I )}  = \det{- 2 i \lambda \Gamma + \lambda^2 - K} = 0, 
\end{align}
where we have used the identity ${\det A \det B = \det{A B}}$. 
Making ${\lambda = -\omega}$ gives
\begin{align} \label{a_208}
    & p_{\sqrt \mathcal K} (- \omega ) = \det{ \omega^2 + 2 i \omega \Gamma - K} = 0.
\end{align}
Noting that ${p_{\sqrt \mathcal K} (- \omega )} = 0 \Leftrightarrow  p_{- \sqrt \mathcal K} ( \omega ) = 0 $ results 
\begin{align} \label{a_246}
    & p_{- \sqrt \mathcal K} ( \omega ) = \det{ \omega^2 + 2 i \omega \Gamma - K} = 0, 
\end{align}
so ${-\sqrt \mathcal K}$ has the same eigenvalues as the kernel ${A(\omega)}$ in the ${\omega}$-space representation of \cref{a_133}:
\begin{subequations} \label{a_241}
\begin{align} 
    & u(\omega) = A(\omega) f(\omega) \\
    & A(\omega):=[ \omega^2 + 2 i \omega \Gamma - K ]^{-1}.
\end{align}
\end{subequations}

\end{proof}

Let's now show that ${p_{J\mathcal B} (\lambda) := \det{J \mathcal B -\lambda E} = 0}$ is verified by ${\lambda \in Sp\{\pm i \sqrt \mathcal K\} \cup \{0\}}$, where  ${ J \mathcal B = \begin{bmatrix} 0 & A^{-1} \mathcal K \\ -A & 0 \end{bmatrix}}$ and ${E : = \begin{bmatrix} \text{Id} & 0 \\ 0 & \text{Id} \end{bmatrix}}$.

\begin{proof}
    Using the formula for the determinant of a block matrix (\cref{a_205}) and assuming ${\lambda \neq 0}$, we have 
    \begin{align} \label{a_218}
        & p_{J\mathcal B} (\lambda) = \det{\begin{bmatrix} -\lambda\, \text{Id} & A^{-1} \mathcal K \\ - A & - \lambda\, \text{Id} \end{bmatrix} } 
            = \det{-\lambda\, \text{Id} } \det{ -\lambda\, \text{Id} + A (- \lambda^{-1}\, \text{Id}) A^{-1} \mathcal K } \nonumber\\
        & = \det{ \lambda^2\, \text{Id} + \mathcal K } = \det{ \lambda\, \text{Id} + i \sqrt \mathcal K }\det{ \lambda\, \text{Id} - i \sqrt \mathcal K }  = 0.
    \end{align}
    From the above results we have 
    \begin{align} \label{a_240}
        & Sp\{ J \mathcal B\} \subseteq \pm Sp \big \{ i \sqrt \mathcal K \big\} \cup \{0\}.
    \end{align}
    Since ${Sp \big \{ - \sqrt \mathcal K \big\} \subseteq \big\{ \omega : \det( \omega^2 + 2 i \omega \Gamma - K)=0 \big\} \cup \{0\}}$, we can omit the zero in \cref{a_240} to write ${Sp\{ J \mathcal B\} = \pm Sp \big \{ i \sqrt \mathcal K \big\}}$.

\end{proof}

The preceding results are summarized in the following equations:
\begin{tcolorbox}{Spectral Theorems:}
    \begin{subequations} \label{a_239}
    \begin{align} 
        & Sp \big \{ - \sqrt \mathcal K \big\} \subseteq \big\{ \omega : \det( \omega^2 + 2 i \omega \Gamma - K ) = 0 \big\} \cup \{0\} \\
        & Sp\{ J \mathcal B\} = \pm Sp \big \{ i \sqrt \mathcal K \big\} .
    \end{align}
    \end{subequations}

\end{tcolorbox}

\section{Hamiltonian Formulation in the Extended Variables} \label{a_264}

    For this system, we can propose the Hamiltonian 
    \begin{align} \label{a_172}
        &  H (t) = \frac{1}{2} \pi^\mathrm T A \pi  + \frac{1}{2} x^\mathrm T A^{-1} \mathcal K x + x^\mathrm T A^{-1} F_t 
    \end{align}
    where ${A}$ is the symmetric and invertible matrix that performs the similarity transformation~\cite{taussky1959similarity}
    \begin{align} \label{a_174}
        & \mathcal K = A \mathcal K^\mathrm T A^{-1} . 
    \end{align}
    Defining the canonical vector as ${q := \begin{bmatrix} \pi & x \end{bmatrix}^\mathrm T }$, of ${\dim = 2N}$ with ${N=2n}$, where ${x= \begin{bmatrix} u & v \end{bmatrix}^\mathrm T}$ and ${\pi  := \begin{bmatrix} \pi_u & \pi_v \end{bmatrix}^\mathrm T}$, together with the matrices 
    \begin{align}
        & \mathcal B  :=  \begin{bmatrix} A & 0 \\ 0 & A^{-1} \mathcal K \end{bmatrix} \\
        & \mathcal{C}_t := \begin{bmatrix} 0 & A^{-1} F_t \end{bmatrix}^\mathrm{T} , 
    \end{align}
    we can write \cref{a_172} as 
    \begin{align}
        & H(t) = \frac{1}{2} q^\mathrm T \mathcal B q + q ^\mathrm T \mathcal C_t ,
    \end{align}
    where ${\mathcal B  = \mathcal B ^\mathrm T}$ is time-independent.
The vector components ${q_\xi }$ are operators verifying the Canonical Commutation Relation (CCR)
\begin{align} \label{a_224}
    & \comm{  q_\xi}{ q_\eta} = - i \hbar  J _{\xi \eta} ,\quad \xi,\eta \in \{1,\dots, 2 N\}, 
\end{align}
where  ${ J = \begin{bmatrix} 0 & I \\ - I & 0 \end{bmatrix} }$ is the standard symplectic matrix in ${ \mathbb R^{2 N\times 2 N} }$, ${I}$ the identity in ${ \mathbb R^{N\times N} }$ and ${J^\mathrm T = J ^{-1} = -J }$.

\subsection{The Similarity Transformation} \label{a_139}
 
To find the matrix $A$ symmetric and invertible that achieves the similarity transformation of $\mathcal K$ into $\mathcal K^\mathrm T$, namely~\cite{taussky1959similarity}
 \begin{align} \label{a_140}
     & \mathcal K = A \mathcal K^\mathrm T A^{-1},
 \end{align}
 we can follow the strategy outlined in~\cite{Kaczorek_2005}.
 By the Jordan normal form theorem, ${\forall\, \mathcal K \in \mathbb C ^{n\times n} }$ there exists a non-singular matrix ${P_1 \in \mathbb C ^{n\times n}}$ such that 
\begin{align} \label{a_141}
    & \mathcal K = P_1 J_\mathcal K P_1 ^{-1} \\
    & J_\mathcal K = \diag \begin{bmatrix} J_1 & J_2 & \dots & J_s \end{bmatrix} 
\end{align}
where 
\begin{align} \label{a_142}
    & J_i = \begin{bmatrix} \lambda_i & 1 & 0 &  \dots & 0 \\ 0 & \lambda_i & 1 & \dots & 0 \\
    \vdots & \vdots & \vdots & \ddots &  \vdots \\
    0 & 0 & 0 & \dots & 1 \\
    0 & 0 & 0 & \dots & \lambda_i \end{bmatrix},\, i =1,\, \dots,\, s 
\end{align}
are the Jordan blocks. Let  
\begin{align} \label{a_144}
    & P_{2,i } = \begin{bmatrix} 
        0 & 0 & \dots & 0 & 1 \\
        0 & 0 & \dots & 1 & 0 \\
    \vdots & \vdots & \vdots & \ddots &  \vdots \\
    1 & 0 & \dots & 0 &  0 \end{bmatrix} = P_{2,i} ^{-1} = P_{2,i} ^\mathrm T ,
\end{align}
be the exchange matrix of ${\dim= i^2}$, then it is easy to verify that the following equality holds,
\begin{align} \label{a_203}
    & J_ i ^\mathrm T = P_{2,i} J_i P_{2,i}\\
\label{a_143}
    & J_\mathcal K ^\mathrm T = P_2 J_\mathcal K P_2 , 
\end{align}
where ${ P_2 = \diag\, [ P_{2,1}\,  P_{2,2}\, \dots\, P_{2,s} ] }$ is a block diagonal matrix.
From \cref{a_141,a_144,a_143,a_203} we have 
\begin{align} \label{a_145}
    & \mathcal K^\mathrm T =  (P_1 ^{-1})^\mathrm T J_\mathcal K^\mathrm T P_1 ^\mathrm T = (P_1 ^{-1})^\mathrm T P_2 J_\mathcal K P_2 P_1 ^\mathrm T 
    = (P_1 ^{-1})^\mathrm T P_2 P_1 ^{-1} \mathcal K  P_1 P_2 P_1 ^\mathrm T = A^{-1} \mathcal K A, 
\end{align}
where 
\begin{align} \label{a_146}
    & A := P_1 P_2 P_1 ^\mathrm T
\end{align}
is an invertible and symmetric matrix that achieves the similarity transformation.

\subsection{Ehrenfest Dynamics} \label{a_157}

By imposing the canonical quantization relation
\begin{align} \label{a_155}
    & \comm{  q_\xi}{ q_\eta} = - i \hbar  J _{\xi \eta} ,\quad \xi,\eta \in \{1,\dots, 2 N\}, 
\end{align}
where  ${ J = \begin{bmatrix} 0 & I \\ - I & 0 \end{bmatrix} }$ is the standard symplectic matrix in ${ \mathbb R^{2 N\times 2 N} }$, with  ${I}$ the identity in ${ \mathbb R^{N\times N} }$ and ${J^\mathrm T = J ^{-1} = -J }$, we reinterpret \cref{a_156} (and consequently \cref{a_133}) as the Ehrenfest dynamics for the quantum operators ${q_\xi}$.
This can be seen from the identities
\begin{align} \label{a_187}
    & - \frac{i}{\hbar}\comm{\pi}{H(x,\pi)}  = - \nabla_x H (x,\pi) = - A^{-1} \mathcal K x - A^{-1} F_t \\
    \label{a_188}
    & - \frac{i}{\hbar}\comm{x}{H(x,\pi)} = \nabla_\pi H(x,\pi) = A \pi .
\end{align}
These result from substituting \cref{a_172} for ${H(x,\pi)}$ and the identity~\cite{quintela2024quantum}
\begin{align} \label{a_186}
    & \comm{  q}{H(q)} = - i \hbar  J \nabla H (q) , \quad q = \begin{bmatrix} \pi & x \end{bmatrix}^\mathrm T, 
\end{align}
valid when the Hamiltonian does not couple the coordinate $x$ and momenta $\pi$ operators.
In the Heisenberg picture, \cref{a_187,a_188} can be combined with the equations of motion
\begin{align} \label{a_189}
    & \dot \pi = - \frac{i}{\hbar} \comm{\pi}{H(x,\pi)}  = - A^{-1} \mathcal K x - A^{-1} F_t  \\
    & \dot x = - \frac{i}{\hbar} \comm{x}{H(x,\pi)}  = A \pi .
\end{align}
From the expectation values result the Ehrenfest dynamics
\begin{align} \label{a_190}
    & \dot{ \expval \pi} = - A^{-1} \mathcal K \expval x - A^{-1} F_t  \\
\label{a_191}
    & \dot{ \expval x} = A \expval \pi.
\end{align}
Taking the derivative in \cref{a_191}, and substituting \cref{a_190}, results
\begin{align} \label{a_192}
    & \ddot {\expval x} +  \mathcal K \expval x + F_t  = 0 .
\end{align}

\section{Decomposition of dispersive and absorptive amplitudes}\label{sec:decomposition}

We will derive a simple expression for $S_{\vec{s} k}^\mathcal A$ and $S_{\vec{s} k}^\mathcal D$. We start from the definition of $C_{\vec{s}} (\omega)$ expressed as follows:
\begin{align} \label{a_259}
    & C_{\vec s} (\omega) = \begin{bmatrix} (P^{-1})_{11} & (P^{-1})_{12} \\ (P^{-1})_{21} & (P^{-1})_{22} \end{bmatrix} \begin{bmatrix} f(\omega) \otimes R_{\vec s} & 0 \\ - ( i \omega + 2 \Gamma) (f(\omega) \otimes R_{\vec s}) & 0 \end{bmatrix} \begin{bmatrix} P_{11} & P_{12} \\ P_{21} & P_{22} \end{bmatrix} ,
\end{align}
in which the eigenvector matrix is written in blocks, each one of dimension $n \times n$.
Performing the matrix-matrix multiplications, we obtain
\begin{align} \label{a_262}
    & C_{\vec s} (\omega) = \begin{bmatrix} (P^{-1})_{11} & (P^{-1})_{12} \\ (P^{-1})_{21} & (P^{-1})_{22} \end{bmatrix} \begin{bmatrix} (f(\omega) \otimes R_{\vec s}) P_{11} & (f(\omega) \otimes R_{\vec s}) P_{12}  \\ - ( i \omega + 2 \Gamma) (f(\omega) \otimes R_{\vec s}) P_{11} & - ( i \omega + 2 \Gamma) (f(\omega) \otimes R_{\vec s}) P_{12} \end{bmatrix} \nonumber\\
    & = \begin{bmatrix} \big[ (P^{-1})_{11} - (P^{-1})_{12}\, ( i \omega + 2 \Gamma) \big] \, (f(\omega) \otimes R_{\vec s}) P_{11} & * \\ * & \big[ (P^{-1})_{21} - (P^{-1})_{22}\, ( i \omega + 2 \Gamma)]\, (f(\omega) \otimes R_{\vec s}) P_{12} \end{bmatrix}  ,
\end{align}
where the off-diagonal blocks have not been expressed since they do not enter the definition of the amplitudes. 
The diagonal elements become:
\begin{align} \label{a_261}
    & C_{\vec{s},kk,1} = \big[\big( (P^{-1})_{11} - 2 (P^{-1})_{12} \Gamma \big)\, (f(\omega) \otimes R_{\vec s}) P_{11} \big]_{kk} - i \omega \big[ (P^{-1})_{12} (f(\omega) \otimes R_{\vec s}) P_{11} \big]_{kk} \\
    & C_{\vec{s},kk,2} = \big [\big( (P^{-1})_{21} - 2 (P^{-1})_{22} \Gamma \big)\, (f(\omega) \otimes R_{\vec s}) P_{12} \big]_{kk} - i \omega \big[ (P^{-1})_{22} (f(\omega) \otimes R_{\vec s}) P_{12} \big]_{kk} .
\end{align}
Suppose that the force acting on the polarization sources is a kick pulse. 
In this case, the Fourier transform is constant ($f(\omega) = f$), and $C_{\vec{s},kk} (\omega)$ becomes a linear function of the frequency. We can define the following quantities:
\begin{subequations} \label{a_258}
\begin{align} 
    & I_{k,1} := [\big( (P^{-1})_{11} - 2 (P^{-1})_{12} \Gamma\big) (f(\omega) \otimes R_{\vec s}) P_{11}]_{kk} \label{a_1001}\\
    & I_{k,2} : = [\big( (P^{-1})_{21} - 2 (P^{-1})_{22} \Gamma\big) (f(\omega) \otimes R_{\vec s}) P_{12}]_{kk} \label{a_1002}\\
    & A_{k,1} : = [ (P^{-1})_{12}\, (f(\omega) \otimes R_{\vec s}) P_{11}]_{kk} \label{a_1003}\\ 
    & A_{k,2} : = [ (P^{-1}) _{22}\, (f(\omega) \otimes R_{\vec s}) P_{12}]_{kk} \label{a_1004}.
\end{align}
\end{subequations}
By collecting the block contributions together as ${I := \begin{bmatrix} I_1 & I_2 \end{bmatrix}^\mathrm T }$ and  ${A := \begin{bmatrix} A_1 & A_2 \end{bmatrix}^\mathrm T }$, the absorptive and dispersive amplitudes can be written as:
\begin{align} \label{a_260}
    & S_{\vec{s} k}^\mathcal A (\omega) = \Re[I_{k}] - i\omega \Im[A_{k}] \\
    \label{a_263}
    & S_{\vec{s} k}^\mathcal D (\omega) = \Im[I_{k}] - i\omega \Re[A_{k}] .
\end{align}

\section{von Neumann equation for non-Hermitian Hamiltonians } \label{a_225}

We will show that, if the density matrix is evolved as 
\begin{align} \label{a_226}
    & \rho (t,t_0) =  U (t,t_0) \rho(t_0)  U^{-1} (t,t_0) \\
    \label{a_228}
    & U (t,t_0) = \mathcal T \exp{ -i /\hbar \int _{t_0} ^t  H (\tau ) \dd \tau }, 
\end{align}
where ${H(t)}$ is a non-Hermitian operator, then the density matrix evolves according to the von Neumann equation 
\begin{align} \label{a_227}
    & \dv{}{t}\rho(t) = - \frac{i}{\hbar} \comm{H(t)}{\rho (t)}.
\end{align}
\begin{proof}

    The evolution operator satisfies the Schrödinger-like equation 
    \begin{align} \label{a_230}
        & \dv{}{t} U(t,t_0) = - \frac{i}{\hbar} H(t) U(t,t_0), 
    \end{align}
    which after formal integration results
    \begin{align} \label{a_231}
        & U(t,t_0) = \mathcal T \exp{ -i /\hbar \int _{t_0} ^t  H (\tau ) \dd \tau }.
    \end{align}
    By definition it verifies
    \begin{align} \label{a_229}
        & U (t,t_0) U (t_0,t)  = U (t,t_0) U^{-1} (t,t_0) = I ,
    \end{align}
    from which we have
    \begin{align} \label{a_232}
        & \dv{}{t}[U (t,t_0) U^{-1} (t,t_0)] = 0 \Rightarrow \dv{}{t}U^{-1} (t,t_0) = - U^{-1} (t,t_0) \dv{}{t}U (t,t_0) U^{-1} (t,t_0).
    \end{align}
    Substituting \cref{a_230} we have 
    \begin{align} \label{a_233}
        & \dv{}{t}U^{-1} (t,t_0) = \frac{i}{\hbar} U^{-1} (t,t_0) H(t).
    \end{align}

    Taking the derivative in \cref{a_226} 
    \begin{align} \label{a_234}
        & \dv{}{t} \rho (t,t_0) =  \dv{}{t} \Big[ U (t,t_0) \rho(t_0)  U^{-1} (t,t_0) \Big] 
        = \Big( \dv{}{t} U (t,t_0) \Big) \rho(t_0)  U^{-1} (t,t_0) + U (t,t_0) \rho(t_0)  \Big( \dv{}{t} U^{-1} (t,t_0)\Big) .
    \end{align}
    Substituting \cref{a_230,a_233} gives
    \begin{align} \label{a_235}
        & \dv{}{t} \rho (t,t_0) = - \frac{i}{\hbar} H(t) U(t,t_0) \rho(t_0)  U^{-1} (t,t_0) + \frac{i}{\hbar} U (t,t_0) \rho(t_0)U^{-1} (t,t_0) H(t) =- \frac{i}{\hbar} \comm{H(t)}{\rho(t)}.
    \end{align}
\end{proof}

\section{The Delta Function} \label{a_167}

In what follows, a function of a matrix $f(M)$ is nothing but an ordinary function $f(x)$, $x \in  \mathbb R$, where the variable $x$ is replaced by the matrix, $f (M) = f(x)\Big|_{x=M}$. 
This is clear when $f(x)$ can be expressed as a power series, where the replacement operation $x \xrightarrow[]{repl.by} M$ is well-defined.

\begin{enumerate}[A.]
    \item  
        \begin{align} \label{a_51}
            & \delta (\Omega) = \frac{1}{2\pi} \int_{\mathbb R} \dd s\, e^{i s \Omega } = \delta (- \Omega) .
        \end{align}
    \item 
        \begin{align} \label{a_52}
            & \int_\mathbb R \dd x\, \delta (x E - \Omega ) f(x) 
              = \int_\mathbb R \dd x\, \frac{1}{2\pi} \int_\mathbb R  \dd s\, e^{i s ( x E - \Omega ) }  f(x)  = \frac{1}{2\pi} \int_\mathbb R  \dd s\, e^{-i s \Omega } \Big( \int_\mathbb R \dd x\,  e^{i s x E } f(x) \Big)  \nonumber\\
            & = \frac{1}{2\pi} \int_\mathbb R  \dd s\, e^{-i s \Omega }  \tilde f(s) 
              = \frac{1}{2\pi} \int_\mathbb R  \dd s\, e^{-i s x } \tilde f(s)  \Big|_{x=\Omega }  
              =  f(x) \Big|_{x=\Omega } = f(\Omega ) ,
        \end{align}
        provided $f(x)$ has a power series representation. 
        Then we have the rule 
        \begin{align} \label{a_53}
            & \int_\mathbb R \dd x\, \delta (x E - \Omega ) f(x) =  f(\Omega ) .
        \end{align}
    \item Let's now show the identity 
        \begin{align} \label{a_54}
            & \delta (M- \omega E) f(M) =  \delta (M- \omega E) f(\omega ) ,
        \end{align}
        Using \cref{a_53}, the \textit{l.h.s} of \cref{a_54} can be written as 
        \begin{align} \label{a_55}
            & l.h.s = \int _\mathbb R \dd \omega ' \delta ( M - \omega E ) \delta (M -\omega' E ) f (\omega ') .
        \end{align}
        Using the identity (proven below)
        \begin{align} \label{m_81}
            & \delta (M - y E ) \delta (M - x E) = \delta (M - y E) \delta (y - x ) 
        \end{align}
        results in
        \begin{align} \label{a_56}
            & l.h.s = \int _\mathbb R \dd \omega '  \delta (M - \omega E ) \delta (\omega - \omega' ) f (\omega ') =  \delta (M - \omega E ) f(\omega) .
        \end{align}

    \item We now show \Cref{m_81}. Consider the integral 
        \begin{align} \label{a_57}
            & \int_{\mathbb R^2} \dd x\, \dd y\, \delta (M - y E ) \delta (M - x E) f(x,y) 
              = \int_\mathbb R \dd x\, \delta (M - x E )  f(x,M) =  f(M,M) \nonumber\\ 
            & = \int_{\mathbb R^2} \dd x\, \dd y\, \delta (M - y E) \delta (y E - x E ) f(x,y).
        \end{align}
        Subtracting the extremes we have
        \begin{align} \label{a_58}
            & \int_{\mathbb R^2} \dd x\, \dd y\, I (x,y) f(x,y)  = 0\nonumber\\
            & I (x,y) \equiv \delta (M - y E)  \Big( \delta (y E - x E ) - \delta (M - x E)  \Big) .
        \end{align}
        Let's now see that ${ I (x,y) = 0,\, \forall x,y}$. Note that
        \begin{align} \label{a_59}
            & I(x,y) = \begin{cases} \delta (0) \cdot 0, & y E = M \\ 0 \cdot \Big( \delta (y E - x E ) - \delta (M - x E)  \Big) , & y E \neq M \end{cases} .
        \end{align}
        Using 
        \begin{align} \label{a_60}
            &  \delta (\Omega ) = i \lim_{\epsilon \to 0^+ } \Big((\Omega + i \epsilon E)^{-1} - (\Omega - i \epsilon E)^{-1}\Big) 
        \end{align}
        we see that 
        \begin{align} \label{a_61}
            &  \delta (\Omega ) \cdot 0 = i \lim_{\epsilon \to 0^+ } \Big((\Omega + i \epsilon E)^{-1} - (\Omega - i \epsilon E)^{-1}\Big) \cdot 0  \nonumber\\
            & = \lim_{\epsilon \to 0^+ } 0 = 0, 
        \end{align}
        so $I(x,y)=0$. Now the identity ${\delta (y E  - x E ) = E \delta (y - x ) E }$ completes the proof.
\end{enumerate}

\section{Field interacting with the Auxiliary Variables} \label{a_168}

\begin{align} \label{a_169}
    & H_\mathrm{int} (t) = H_1(t) + H_2(t) = \sum_{\nu  = 1 }^n x_\nu h_\nu (t)  + x_{n + \nu} g_\nu (t)  
    = x^\mathrm T \begin{bmatrix} h(t) \\ g(t) \end{bmatrix} .
\end{align}
where ${h_\nu (t)}$ is an arbitrary time-dependent function, that might be of the form
\begin{align} \label{a_170}
    & h_\nu (t) \equiv - \sum_{j=1}^3 R_{j \nu } \int_{\mathbb R ^3 } \dd r\, \mathcal E_j  (r, t) G (r; R_\nu, \Sigma_\nu ) .
\end{align}
Comparing with \cref{a_173,a_172}, and indicating the time-dependence by the subscript $t$, we have 
\begin{align} \label{a_171}
    & F_t = A \begin{bmatrix} h_t \\ g_t \end{bmatrix} \Leftrightarrow 
      \begin{bmatrix} f_t \\ \dot f_t - 2 \Gamma f_t \end{bmatrix} 
      = \begin{bmatrix} A_1 & A_2 \\ A_2^\mathrm T & A_3 \end{bmatrix} \begin{bmatrix} h_t \\ g_t \end{bmatrix}, 
\end{align}
with ${A_1 = A_1^\mathrm T}$, ${A_3 = A_3^\mathrm T}$, and $A$ given by \cref{a_174}.
Separating the vector components, 
\begin{align} \label{a_175}
    & A_1 h_t + A_2 g_t = f_t \\
    \label{a_176}
    & A_2^\mathrm T h_t + A_3 g_t = \dot f_t -2 \Gamma f_t .
\end{align}
Substituting \cref{a_175} into \cref{a_176} we have 
\begin{align} \label{a_177}
    & A_2^\mathrm T h_t + A_3 g_t =  A_1 \dot h_t + A_2 \dot g_t -2 \Gamma (A_1 h_t + A_2 g_t ) 
\end{align}
or, equivalently, 
\begin{align} \label{a_178}
    & A_2 \dot g_t -( A_3 + 2 \Gamma A_2 ) g_t = (A_2^\mathrm T + 2 \Gamma A_1 ) h_t - A_1 \dot h_t .
\end{align}
In the $\omega$ space, 
\begin{align} \label{a_179}
    & -i \omega A_2 \tilde g(\omega) -( A_3 + 2 \Gamma A_2 ) \tilde  g(\omega) = (A_2^\mathrm T + 2 \Gamma A_1 )\tilde  h(\omega) + i\omega  A_1 \tilde h (\omega) .
\end{align}
Grouping terms and isolating ${\tilde g (\omega)}$ yields 
\begin{align} \label{a_180}
    & \tilde g (\omega ) = \tilde {\mathcal L} (\omega ) \tilde h (\omega )\\
    \label{a_181}
    & \tilde{ \mathcal L}(\omega ) := - \big[ A_3 + (i \omega + 2 \Gamma ) A_2 \big] ^{-1} \big[ A_2^\mathrm T +  (i \omega + 2 \Gamma ) A_1 \big] .
\end{align}

\subsection{Formal integration} \label{a_251}

Taking the Fourier Transform (\cref{a_93}) in \cref{a_180} we have 
\begin{align} \label{a_250}
    & g_i (t) = \sum_{k=1}^{n} \mathcal \mathscr F \{ \tilde{\mathcal L}_{ik} (\omega) \tilde h_k (\omega) \} = \sum_{k=1}^{n} \mathscr F \{ \tilde{\mathcal L}_{ik} (\omega) \tilde h_k (\omega) \}
    = \sum_{k=1}^{n} \frac{1}{2\pi} \int_{\mathbb R} \dd \omega\,  \exp{ -i \omega t} \tilde{\mathcal L}_{ik} (\omega) \tilde h_k (\omega) \nonumber\\
    & = \sum_{k=1}^{n} \frac{1}{2\pi} \int_{\mathbb R} \dd \omega\,  \exp{ -i \omega t} \tilde{ \mathcal L}_{ik} (\omega) \Big[ \int_{\mathbb R} \dd \tau\, \exp{ i \omega \tau} h_k (\tau) \Big]  
    = \sum_{k=1}^{n} \int_{\mathbb R} \dd \tau\, h_k (\tau) \Big[ \frac{1}{2\pi} \int_{\mathbb R} \dd \omega\,  \exp{ -i \omega (t -\tau) } \tilde{ \mathcal L}_{ik} (\omega)  \Big] \nonumber\\
    & =  \sum_{k=1}^{n} \int_{\mathbb R} \dd \tau\, h_k (\tau) \mathcal L_{ik} (t-\tau) 
    =  \sum_{k=1}^{n} \int_{\mathbb R} \dd \tau\, \mathcal L_{ik} (t) h_k (t -\tau) 
    = \sum_{k=1}^{n} \mathcal L_{ik} (t) \ast h_k (t) \equiv \mathcal L(t) \circledast h(t).
\end{align}
where we have used \cref{a_93,a_102} and the commutativity of the convolution.
Finally, we have introduced the symbol ${\circledast}$ to indicate the convolution of a matrix and a vector.

\section{Quantities in the \texorpdfstring{${k, \omega}$}{k,w} space} \label{a_150}

\subsection{Polarization} \label{a_84}

Defining ${ \Lambda_t ^\pm \equiv \exp{\pm J \mathcal B t} }$  we have
\begin{align} \label{a_104}
    & \mathscr F \{ \Lambda_t ^{\pm } \} (\omega ) = \int_{\mathbb R} \dd t\, \exp{ i ( \omega E \mp i J \mathcal  B) t}  = 2 \pi \delta ( \omega E \mp i J \mathcal  B) .
\end{align}

Taking the Fourier transform of 
\begin{align} \label{a_184}
    & \expval{  q} _t =\Lambda ^{-1}_t ( \expval{   q}_0 - \Delta _t) ,
\end{align}
we have 
\begin{subequations} \label{a_236}
\begin{align} \label{a_107}
    & \expval q (\omega)  = \mathscr F \{ \Lambda_t ^{-1} \expval q_0 - \Lambda_t ^{-1} \Delta_t  \} (\omega )  \\
    & = 2 \pi \delta ( \omega E + i J \mathcal  B) \expval q_0 - \frac{1}{2\pi} (\Lambda_t ^{-1}) (\omega) \circledast \Delta (\omega) \\
    & = 2 \pi \delta (\omega E + i J \mathcal B) \expval q_0 - i ( \omega E + i J \mathcal B )^{-1} J \mathcal C (\omega) ,
\end{align}
\end{subequations}
    where we have used the convolution theorem for matrices (see the proof below)
    \begin{align} \label{a_193}
        & \mathscr F \{ A(t) B(t) \} (\omega) = \frac{1}{2\pi} A(\omega) \circledast B (\omega) ,
    \end{align}
    where ${ [A \circledast B ]_{ij} (\omega) \equiv \sum_k (A_{ik} \ast B_{kj} ) (\omega) }$ is the matrix convolution, and the identity (see the proof below)
    \begin{align} \label{a_119}
        & \Lambda^{-1}(\omega) \circledast \Delta (\omega) = 2 \pi i (\omega E + i J \mathcal B)^{-1} J \mathcal C(\omega ) .
    \end{align}
\begin{proof}

    The Fourier transform of the product of functions reads
\begin{align} \label{m_115}
    & \mathscr  F \{ f(t) g (t) \} ( \omega ) 
    =  \int_{\mathbb R} \dd t\, \exp{ i \omega t} f(t) g (t)  
    = \int_{\mathbb R} \dd t\, \exp{ i \omega t} \frac{1}{2\pi} \int_{\mathbb R} \dd \omega'\,  \exp{ -i \omega' t} \tilde f(\omega ') g(t) \nonumber\\
    & = \frac{1}{2\pi} \int_{\mathbb R} \dd \omega'\,\tilde f(\omega ')  \int_{\mathbb R} \dd t\, \exp{ i ( \omega - \omega') t} g(t) 
    = \frac{1}{2\pi} \int_{\mathbb R} \dd \omega'\,\tilde f(\omega ')  \tilde g ( \omega - \omega') 
    = \frac{1}{2\pi} \tilde f (\omega) \ast \tilde g (\omega) .
\end{align}
Then results the convolution theorem in the $\omega$ space,
\begin{align} \label{a_103}
    & \mathscr  F \{ f(t) g (t) \} ( \omega ) = \frac{1}{2\pi} \tilde f (\omega) \ast \tilde g (\omega) .
\end{align}

    The convolution for scalars is defined as 
    \begin{align} \label{a_126}
        & f \ast g (\omega) = \int_{\mathbb R} \dd \omega' f(\omega -\omega ') g(\omega ') . 
    \end{align}
    This is extended to matrices as follows: From the definition of ${\circledast}$ we have 
    \begin{align} \label{a_131}
        & (A \circledast B)_{ij } (\omega) \equiv \sum_k (A_{ik} \ast B_{kj} ) (\omega)  
        = \sum_k  \int_{\mathbb R} \dd \omega' A_{ik} (\omega -\omega ') B_{kj} (\omega ')  \nonumber\\
        & =  \int_{\mathbb R} \dd \omega' \sum_k A_{ik} (\omega -\omega ') B_{kj} (\omega ')  
        = \Big( \int_{\mathbb R} \dd \omega' A (\omega -\omega ') B (\omega ') \Big)_{ij} .
    \end{align}
    For matrices, the convolution theorem reads
    \begin{align} \label{a_194}
        & \mathscr F \{ A (t) B(t)\}_{ij} (\omega ) = \mathscr F \Big \{ \sum_k A_{i k} (t) B_{k j}(t) \Big \} (\omega ) 
        = \sum_k  \mathscr F \{ A_{i k} (t) B_{k j}(t)\} (\omega ) \nonumber\\
        & = \frac{1}{2\pi } \sum_k A_{i k} (\omega) \ast B_{k j} ( \omega) = \frac{1}{2\pi} \Big( A(\omega) \circledast B (\omega) \Big) _{ij}, 
    \end{align}
    where we have used linearity of the Fourier transform. This proves \cref{a_193}.

\end{proof}
\begin{proof}
    Let's now prove \cref{a_119}. Using ${\mathscr F \{ \dot \Delta _t \} (\omega) = - i \omega \Delta (\omega )}$ and the convolution theorem, we can write ${\dot \Delta_t  = \Lambda_t J \mathcal C_t}$ in the ${\omega}$ space as 
\begin{align} \label{a_105}
    & - i \omega \Delta (\omega ) = \mathscr F \{ \Lambda_t J \mathcal C_t \} (\omega)
    =  \frac{1}{2\pi } \Lambda (\omega ) \circledast J \mathcal C (\omega)  = \delta ( \omega E - i J \mathcal  B) \circledast J \mathcal  C (\omega) , 
\end{align}
    where we have used \cref{a_104}.
    From the convolution for matrices, \cref{a_131}, we have 
    \begin{align} \label{a_127}
        & \delta (\omega E - i J \mathcal B) \circledast J \mathcal C (\omega) 
        = \int_{\mathbb R} \dd \omega' \delta (\omega E - i J \mathcal B -\omega ' E ) J \mathcal C (\omega')  
        = \int_{\mathbb R} \dd \omega' \delta (( \omega E - i J \mathcal B) -\omega ' E ) J \mathcal C (\omega') \nonumber\\
        & = \int_{\mathbb R} \dd \omega' \delta (( \omega E - i J \mathcal B) -\omega ' E ) J \mathcal C ( \omega E - i J \mathcal B)  = \Big(  \int_{\mathbb R} \dd \omega' \delta (( \omega E - i J \mathcal B) -\omega ' E ) \Big) J \mathcal C ( \omega E - i J \mathcal B) ,
    \end{align}
    where we have used the delta function identity \cref{a_54}.
    Using \cref{a_53} we have 
    \begin{align} \label{a_128}
        & \int_{\mathbb R} \dd \omega' \delta (( \omega E - i J \mathcal B) -\omega ' E ) =  E ,
    \end{align}
    so
    \begin{align} \label{a_129}
        & \delta (\omega E - i J \mathcal B) \circledast J \mathcal C (\omega) = J \mathcal C ( \omega E - i J \mathcal B)  .
    \end{align}
    Then we have 
    \begin{align} \label{a_195}
        & \Delta (\omega) =  i \omega^{-1} J \mathcal C ( \omega E - i J \mathcal B)  .
    \end{align}
    Now we use this result to compute the convolution
    \begin{align} \label{a_130}
        & \Lambda^{-1}(\omega) \circledast \Delta (\omega)  = 2 \pi \delta (\omega E + i J \mathcal B) \circledast \Big( i \omega^{-1} J \mathcal C ( \omega E - i J \mathcal B) \Big)  =  2 \pi i (\omega E + i J \mathcal B)^{-1} J \mathcal C ( \omega ) 
    \end{align}
    by the same token as \cref{a_129}. This proves \cref{a_119}.

\end{proof}

\subsection{Spatial Distribution and the Interaction Term } \label{a_98}

The Fourier transform of the Gaussian function
\begin{align} \label{a_182}
    & G (r; R_\alpha, \Sigma_\alpha ) = |\det{2 \pi \Sigma_\alpha }|^{-\frac{1}{2}} \exp{ -\frac{1}{2} (r -R_\alpha )^\mathrm T \Sigma_\alpha ^{-1} (r -R_\alpha )}
\end{align}

to the $k$ space, \cref{a_101}, reads
\begin{align} \label{a_108}
    & \tilde G (k; R_\alpha, \Sigma_\alpha ) =\int_{\mathbb R^3 } \dd r\, \exp{ - i k^\mathrm T r} G (r; R_\alpha, \Sigma_\alpha ) \nonumber\\
    & = |\det{2 \pi \Sigma_\alpha }|^{-\frac{1}{2}}\int_{\mathbb R^3 } \dd r\, \exp{ -\frac{1}{2} (r -R_\alpha )^\mathrm T \Sigma_\alpha ^{-1} (r -R_\alpha ) - i k^\mathrm T r} \nonumber\\
    & = |\det{2 \pi \Sigma_\alpha }|^{-\frac{1}{2}}\int_{\mathbb R^3 } \dd x\, \exp{ -\frac{1}{2} x^\mathrm T \Sigma_\alpha ^{-1} x - i k^\mathrm T x - i k^\mathrm T R_\alpha } .
\end{align}
Using the Gaussian integral
\begin{align} \label{a_92}
    & |\det{2 \pi M}| ^{-\frac{1}{2}} \int_{\mathbb R^3} \dd x \exp{ -\frac{1}{2} x^\mathrm T M^{-1} x  \pm b^\mathrm T x + c }  = \exp{\frac{1}{2} b^\mathrm T M b + c }, 
\end{align}
we have 
\begin{align} \label{a_110}
    & \tilde G (k; R_\alpha, \Sigma_\alpha ) = \exp{ -i k^\mathrm T R_\alpha - \frac{1}{2} k^\mathrm T \Sigma_\alpha k } .
\end{align}

For the interaction term ${h_\beta (t)}$ we have 
\begin{align} \label{a_99}
    & h_\beta (t) = - \sum_{j=1}^3 R_{j \beta } \int_{\mathbb R ^3 } \dd r\, \mathcal E_j  (r, t) G (r; R_\beta, \Sigma_\beta ) \\
    & = - \sum_{j=1}^3 R_{j \beta } \int_{\mathbb R ^3 } \dd r\, \frac{1}{(2\pi)^4} \int_{\mathbb R} \dd \omega\, \int_{\mathbb R ^3 } \dd k\, \exp{- i \omega t + i k^\mathrm T r } \tilde{ \mathcal E}_j  (k, \omega ) G (r; R_\beta, \Sigma_\beta ) \nonumber\\
    & = \frac{1}{(2\pi)} \int_{\mathbb R} \dd \omega\, \exp{- i \omega t}  \Bigg [ - \frac{1}{(2\pi)^3} \sum_{j=1}^3 R_{j \beta } \int_{\mathbb R ^3 } \dd k\, \tilde{ \mathcal E}_j  (k, \omega ) \int_{\mathbb R ^3 } \dd r\, \exp{i k^\mathrm T r }  G (r; R_\beta, \Sigma_\beta ) \Bigg]  \nonumber\\
    & = \frac{1}{(2\pi)} \int_{\mathbb R} \dd \omega\, \exp{- i \omega t}  \Bigg [ - \frac{1}{(2\pi)^3} \sum_{j=1}^3 R_{j \beta } \int_{\mathbb R ^3 } \dd k\, \tilde{ \mathcal E}_j  (k, \omega ) \tilde G (- k; R_\beta, \Sigma_\beta ) \Bigg]  ,
\end{align}
where we have written the electric field using \cref{a_95}, and taken the Fourier transform \cref{a_101} for ${G (r; R_\beta, \Sigma_\beta )}$. 
From the Fourier transform~\cref{a_102} we have 
\begin{align} \label{a_100}
    & h_\beta (\omega ) = - \frac{1}{(2\pi)^3} \sum_{j=1}^3 R_{j \beta } \int_{\mathbb R ^3 } \dd k\, \tilde{ \mathcal E}_j  (k, \omega ) \tilde G (-k; R_\beta, \Sigma_\beta ) .
\end{align}

\subsection{List of Fourier Transforms} \label{a_97}

The Fourier transform in the $\omega$ space reads~\cite{mukamel1995principles}
\begin{align} \label{a_93}
    & F (t) = \frac{1}{2\pi} \int_{\mathbb R} \dd \omega\,  \exp{ -i \omega t} \tilde F (\omega )\\
\label{a_102}
    & \tilde F (\omega ) =  \int_{\mathbb R} \dd t\, \exp{ i \omega t} F (t) .
\end{align}
In the $k$ space,
\begin{align} \label{a_94}
    & V (r) = \frac{1}{(2\pi)^3 }\int_{\mathbb R^3 } \dd k\,  \exp{  i k^\mathrm T r} \tilde V (k)  \\
\label{a_101}
    & \tilde V (k) = \int_{\mathbb R^3 } \dd r\, \exp{ - i k^\mathrm T r} V (r) .
\end{align}
In the ${k,\, \omega}$ space 
\begin{align} \label{a_95}
    & \mathcal E (r,t) = \frac{1}{(2\pi)^4 } \int_{\mathbb R } \dd \omega \int_{\mathbb R^3 } \dd k\, \exp{ -i \omega t + i k^\mathrm T r} \tilde{\mathcal E} (k, \omega )  \\
    \label{a_183}
    & \tilde{ \mathcal E} (k,\omega) = \int_{\mathbb R } \dd t \int_{\mathbb R^3 } \dd r\, \exp{ i \omega t - i k^\mathrm T r} \mathcal E (r, t )  .
\end{align}
From above we have the Fourier transforms of the Dirac delta function,
\begin{align} \label{a_96}
    & \delta (\omega ) = \frac{1}{(2\pi) } \int_{\mathbb R } \dd t \, \exp{ i \omega t}  \\
    & \delta (k) = \frac{1}{(2\pi)^3 } \int_{\mathbb R^3 } \dd r\, \exp{ - i k^\mathrm T r} \\
    & \delta (k,\omega ) = \frac{1}{(2\pi)^4 } \int_{\mathbb R } \dd t \int_{\mathbb R^3 } \dd r\, \exp{ i \omega t - i k^\mathrm T r} .
\end{align}

\end{document}